\documentclass[onecolumn]{aastex631}
\usepackage{subfigure}
\usepackage{color}
\usepackage{graphicx}
\usepackage{amsmath}
\usepackage{csvsimple}
\usepackage{mathtools}
\usepackage{enumitem}
\usepackage{graphicx}
\usepackage{mathrsfs}
\usepackage{booktabs}
\usepackage{rotating}
\usepackage{float}
\usepackage{morefloats}
\usepackage{etoolbox}
\usepackage{multirow}
\usepackage{esint}
\usepackage{tikz}
\usepackage{xspace}
\usepackage{wrapfig}
\usepackage{soul}
\AtBeginEnvironment{quote}{\par\singlespacing}

\definecolor{bcolor}{RGB}{0, 51, 153}
\definecolor{gcolor}{RGB}{10, 100, 10}
\definecolor{dblue}{RGB}{100, 50, 200}

\newcommand{\code}[1]{{\texttt{#1}}}

\newcommand{\porb}{P_{\rm orb}}

\newcommand{\kepler}{{\it Kepler}\xspace}

\newcommand{\tess}{{\it TESS}\xspace}
\newcommand{\ktwo}{{\it K2}\xspace}
\newcommand{\gaia}{{\it Gaia}\xspace}

\newcommand{\lk}{{\code{lightkurve}}\xspace}

\shorttitle{tidal synchronization}

\begin{document}

\title{\sc Tidal Synchronization of TESS Eclipsing Binaries}

\author[0009-0000-4365-0204]{Marshall Hobson-Ritz}
\affiliation{Department of Astronomy, University of Washington, 3910 15th Avenue NE, Seattle, WA 98195, USA}
\author[0000-0002-7961-6881]{Jessica Birky}
\affiliation{Department of Astronomy, University of Washington, 3910 15th Avenue NE, Seattle, WA 98195, USA}
\author{Leah Peterson}
\affiliation{Department of Astronomy, University of Washington, 3910 15th Avenue NE, Seattle, WA 98195, USA}
\author[0009-0009-3547-9326]{Peter Gwartney}
\affiliation{Department of Physics and Astronomy, University of Alabama, Tuscaloosa, AL 35487-0324, USA}
\author{Rachel Wong}
\affiliation{Department of Astronomy, University of Washington, 3910 15th Avenue NE, Seattle, WA 98195, USA}
\author[0009-0008-5172-5307]{John Delker}
\affiliation{Department of Astronomy, University of Washington, 3910 15th Avenue NE, Seattle, WA 98195, USA}
\author[0000-0001-5253-1987]{Tyler Gordon}
\affiliation{Department of Astronomy, University of California, Santa Cruz, 1156 High Street, Santa Cruz, CA 95064, USA}
\author[0009-0004-8402-9608]{Samantha Gilbert}
\affiliation{Department of Astronomy, University of Washington, 3910 15th Avenue NE, Seattle, WA 98195, USA}
\author[0000-0002-0637-835X]{James R. A. Davenport}
\affiliation{Department of Astronomy, University of Washington, 3910 15th Avenue NE, Seattle, WA 98195, USA} 
\author[0000-0001-6487-5445]{Rory Barnes}
\affiliation{Department of Astronomy, University of Washington, 3910 15th Avenue NE, Seattle, WA 98195, USA} 

\begin{abstract}
Tidal synchronization plays a fundamental role in the evolution of binary star systems. However, key details such as the timescale of synchronization, efficiency of tidal dissipation, rotational period, and dependence on stellar mass are not well constrained. We present a catalog of rotation periods, orbital periods, and eccentricities from eclipsing binaries (EBs) that can be used to study the role of tides in the rotational evolution of low-mass dwarf (FGKM spectral type) binaries. This study presents the largest catalog of EB orbital and rotational periods ($P_{\rm orb}$ and $P_{\rm rot}$) measured from the \textsl{Transiting Exoplanet Satellite Survey} (\tess).
We first classify 4584 light curves from the \tess\ Eclipsing Binary Catalog according to out-of-eclipse  stellar variability type: starspot modulation, ellipsoidal variability, non-periodic variability, and ``other" variability (e.g. pulsations). We then manually validate each light curve classification, resulting in a sample of 1039 candidates with 584 high-confidence EBs that exhibit detectable star-spot modulation. From there, we measure and compare the rotation period of each starspot-modulated EB using three methods: Lomb-Scargle periodograms, autocorrelation function, and phase dispersion minimization. We find that our period distributions are consistent with previous work that used a sample of 816 starspot EBs from \kepler to identify two populations: a synchronous population (with $P_{\rm orb} \approx P_{\rm rot}$) and a subsynchronous population (with 8$P_{\rm orb} \approx 7P_{\rm rot}$). 
Using Bayesian model comparison, we find that a bimodal distribution is a significantly better fit than a unimodal distribution for \kepler and \tess samples, both individually or combined, confirming that the subsynchronous population is statistically significant. \\
\end{abstract}

\section{Introduction} \label{sec:intro}

More than half of all stars are in binary star systems \citep{duquennoy_multiplicity_1991,raghavan_survey_2010,duchene_stellar_2013}, making them one of the most important objects in astrophysics. The orbits of binary stars enable direct constraints on fundamental stellar parameters, including masses and radii \citep{torres_eclipsing_2018, matson_fundamental_2016}, as well as tidal interactions \citep{hut_tidal_1981,meibom_robust_2005,meibom_observational_2006}. In close systems, binary orbits evolve under the influence of tidal forces, as well as stellar evolution and magnetic braking \citep{witte_orbital_2002,repetto_coupled_2014,penev_poet_2014,bolmont_effect_2016,song_close_2018,fleming_rotation_2019,zanazzi_tidal_2021,Bouchet86}. 
Similar to single stars, each star loses rotational angular momentum due to magnetic breaking amid magnetized stellar winds and mass loss \citep{skumanich_time_1972,matt_mass-dependence_2015}. 
Tidal forces drive an exchange of angular momentum between the orbital and rotational motions of the stars \citep{meibom_observational_2006}. 
As a result, tides dissipate orbital energy as heat, causing systems to evolve toward a state of dynamical equilibrium in which stellar rotation is coplanar and synchronized with the circular orbit \citep{counselman_outcomes_1973,hut_stability_1980}. 
The rate at which binaries approach tidal equilibrium strongly depends on the system's orbital separation, eccentricity, obliquity, rotational speed, and stellar mass ratio \citep{hut_tidal_1981,ferraz-mello_tidal_2008,leconte_is_2010}. 
Therefore, constraining the process of tidal evolution in binary stars, particularly tidal synchronization, requires precise measurements of stellar rotation periods.

Prior to the era of time-domain surveys for measuring stellar variability \cite[e.g.,][]{borucki_kepler_2010,howell_k2_2014,ricker_transiting_2014,chen_zwicky_2020,jayasinghe_asas-sn_2018}, rotational velocities of stars were primarily obtained through spectroscopy, which measures the rotational broadening observed in spectroscopic absorption lines \citep{carroll_rotational_1933,gray_observation_1976,weise_rotational_2010}.
However, constraints from spectroscopy are limited to measuring only the rotational velocity times the projected line-of-sight angle ($v\sin i$), which depends on the orbit inclination ($i$) and stellar radius, which are difficult to constrain without further observations. Furthermore, spectroscopic rotational velocities require fitting or convolving a template atmospheric model to the spectra, making these constraints dependent on many modeling assumptions and complete line lists \citep{gustafsson_grid_2008}.

In contrast to the spectroscopic method, light curve photometry allows for direct measurements of rotation periods from the quasiperiodic variation in starspots over time \citep{aigrain_simple_2012}. 
Surveys including \kepler \citep{borucki_kepler_2010}, \ktwo \citep{howell_k2_2014}, and \textsl{Transiting Exoplanet Satellite Survey} \citep[\tess;][]{ricker_transiting_2014} have vastly expanded the number of stellar rotation measurements, allowing an understanding of how the dynamical evolution of a star depends on stellar properties
\citep{mcquillan_rotation_2014,gordon_stellar_2021,claytor_recovery_2022,angus_inferring_2018,angus_exploring_2020}.
Furthermore, time domain surveys such as \kepler and \ktwo have led to the discovery and characterization of thousands of eclipsing binaries \citep[EBs;][]{prsa_kepler_2011,kirk_keplereclipsing_2016}, including a number of discoveries in open clusters with varying ages \citep{david_k2_2015,david_new_2016,gillen_new_2017,torres_eclipsing_2018}.

In one comprehensive study, \citet[][hereafter L17]{lurie_tidal_2017} used the \kepler\ eclipsing binary catalog \citep[KEBC;][]{prsa_kepler_2011} to investigate the tidal synchronization rates of low-mass dwarf (FGKM spectral type) binary stars. The study continued a long line of synchronization studies of late-type stars with convective envelopes \citep[\textit{e.g.}][]{Giuricin_synchronization,claret_cunha_circularization_1997,meibom_observational_2006,marilli_rotational_2007}. L17 measured rotation periods using star spot modulation to explore tidal synchronization as a function of orbital period, eccentricity, mass ratio, and mass which influence the rates of tidal interaction and evolution. The findings showed: 
\begin{enumerate}
    \item The majority of EBs at orbital periods less than 10 days are tidally synchronized.
    \item There is a transition to higher eccentricity and pseudosynchronization (when eccentric binaries have a net torque of approximately zero, approaching torque equilibrium) at an orbital period of $\sim10$ days.
    \item Synchronization has a stronger dependence on mass ratio when in the very small mass ratio regime.
    \item In the FGK spectral type mass and radius range, there is no discernible dependence on primary star mass.
\end{enumerate}

Alongside these important findings, an unexpected population of EBs was discovered that rotated typically 13\% slower than synchronization. This population had low eccentricities, slightly favored lower mass ratios, and did not show any strong correlation with mass for the FGK primaries. L17 suggested that subsynchronicity is not mass-dependent because no significant correlation was found between primary stellar mass and synchronization in the subpopulation and sample overall. Furthermore, L17 found that the subsynchronous population is nearly circularized and shows no distinction from the rest of the sample in terms of primary star spectral type, leaving no obvious explanation for their existence.

One physical explanation investigated by \cite[][hereafter F19]{fleming_rotation_2019} is whether subsynchronous binaries could result from the competing effects of tidal dissipation and magnetic braking. Under this hypothesis, tidal forces dissipate orbital energy and cause short-period systems to circularize (towards $e\approx0$) and synchronize (towards $P_{\rm orb} \approx P_{\rm rot}$) over time \citep{hut_tidal_1981,ferraz-mello_tidal_2008,leconte_is_2010}. At the same time, magnetic braking causes each star to lose angular momentum due to magnetized stellar winds and mass loss \citep{matt_mass-dependence_2015, breimann}, possibly explaining why some systems near synchronization might rotate slower than expected based on their orbital period.

F19 attempted to reproduce the observed population of \kepler EBs by simulating the evolution of the orbital and rotational dynamics of binaries for a range of plausible initial conditions and tidal parameters. Their model considered the effects of stellar evolution \citep{baraffe_new_2015}, magnetic braking \citep{matt_mass-dependence_2015}, and equilibrium tides \citep{hut_tidal_1981,ferraz-mello_tidal_2008,leconte_is_2010} coupled together under conservation of energy and angular momentum. 
Consistent with the L17 observations, F19 found that shorter than $P_{\rm orb} \lesssim 4$ days, tidal forces dominate over magnetic braking, causing the rotation of the stars to spin up and tidally lock. 
Their work also found that systems with $P_{\rm orb} \gtrsim 4$ days could rotate subsynchronously for $\sim1$ Gyr due to the competing effects of equilibrium tides and magnetic braking. However, while these simulations could reproduce the 1:1 synchronized binaries and generate a wide distribution of subsynchronously rotating binaries, they failed to reproduce this tight overdensity at the 7:8 spin-orbit ratio. F19 found that variations in tidal parameters or initial conditions cannot recreate the L17 distribution.

Another open hypothesis is that subsynchronous binaries could be the result of differential rotation. This explanation, proposed in L17, suggests that latitudinal shear on the surface of these low-mass convective stars could produce star spots at high latitudes that rotate slower than their equatorial period. Because rotation measurements from lightcurves come from the variability produced by starspots on the surface, they could generate observed rotation periods that are systematically slower than the orbital period. Further work needs to be done to explore this hypothesis.

In this work, we investigate two non-physical explanations for the subsynchronous population: the systemic errors due to period-finding algorithm and/or instrumental bias. The L17 investigation was also limited by the faintness of \kepler targets ($K_p < 17$; \citealt{vanderburg_2014}) and the relatively small field of view (100 deg$^2$). \tess is the ideal follow-up to these studies with approximately 85\% sky coverage (2300 deg$^2$), the ability to detect stars in numerous Galactic populations (including those in open clusters) \citep{Fausnaugh_2021}. Note that the EB targets in the \tess field were selected according to eclipse detectability, rather than by apparent magnitude \citep{2014SPIE.9143E..20R}.

We therefore search for the subsynchronous population in a \tess EB sample. If the sub-population persists, it suggests that the phenomenon is not limited to the observational field of \kepler. Additionally, we look for period algorithm bias. We use the same Lomb-Scargle and autocorrelation period finding methods as in L17, but add the additional method of phase-dispersion minimization. We also visually inspect all rotational periods to confirm the automated results.

The remainder of the paper is organized as follows: Section \ref{sec:data} describes our data selection and classification process; Section \ref{sec:methods} describes our procedure for measuring orbital and rotational periods; Section \ref{sec:results} compares the distribution of synchronized binaries to those presented in L17, and finally Section \ref{sec:conclusion} discusses the implications of our results for constraining tidal synchronization as well as future work to be done in this area. \\

\section{Data} \label{sec:data}

We start with the \tess EB catalog \citep[hereafter TEBC;][]{prsa_tess_2021} that contains 4584 eclipsing binaries identified in the first two years (26 sectors) of the \tess survey \citep{ricker_transiting_2014}. The TEBC contains measurements of the orbital period as well as a morphology parameter, which quantifies and effectively reduces the dimensionality of the light curve variability to a continuous variable in the range [0,1], where a lower morphology value corresponds to more detached systems \citep{matijevic_kepler_2012}.

We initially used the \code{lightkurve} package \citep{lightkurve} with the SPOC pipeline to retrieve the first sector of data available for each target. For each target we download 2 minute cadence when available, otherwise we use the 30 minute cadence data. To better constrain the orbital periods of the longer period systems in our sample (6 to 10 days) we download all sectors of data available (each with $\sim28$ day lightcurves) to give at least 4 orbital periods worth of data. Later, we use all sectors for every target in the established subsample, as described in Section \ref{sec:results}.
\begin{figure*}[p]
    \centering
    \includegraphics[scale=0.65]{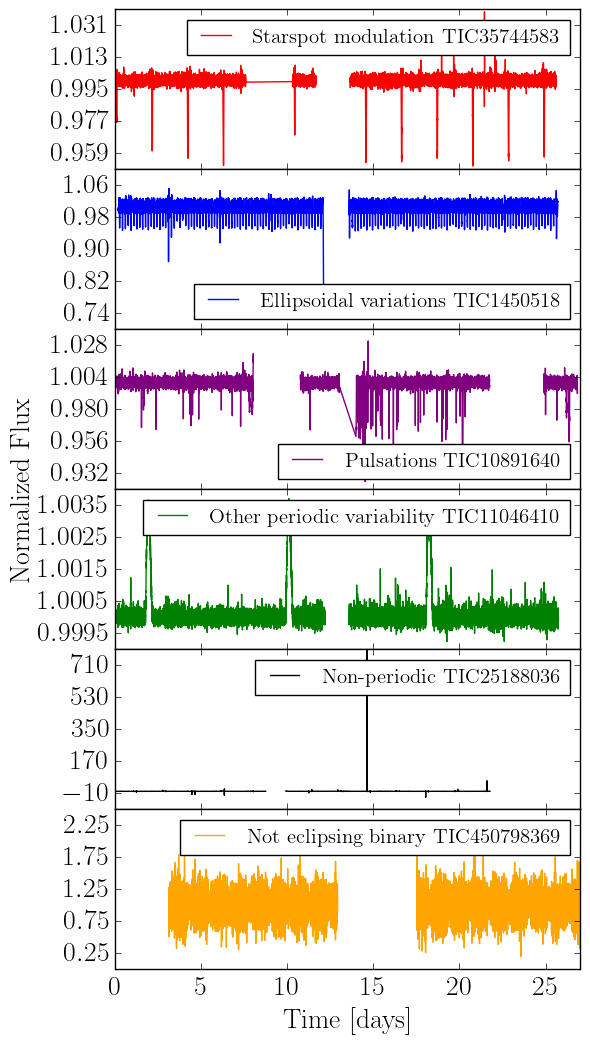}
    \includegraphics[scale=0.65]{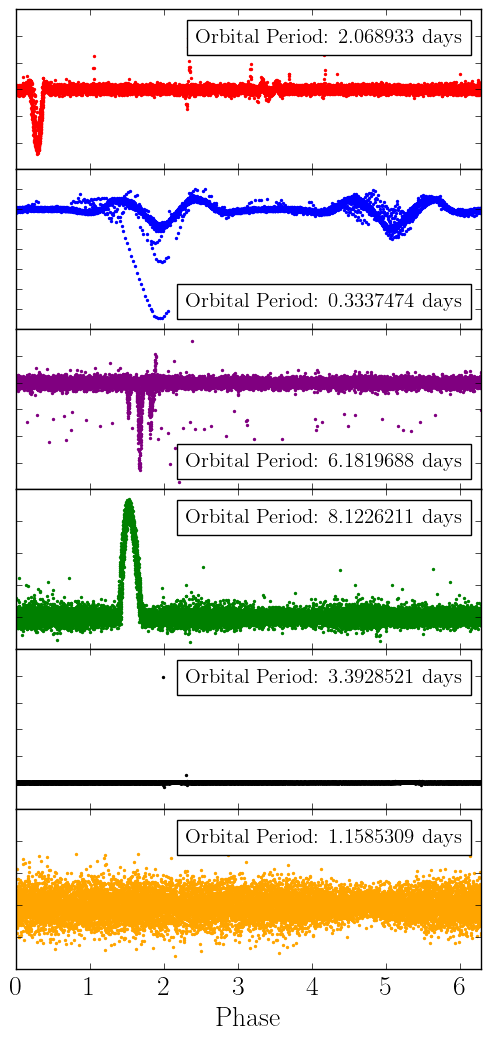}
    \vspace{5pt}
    \caption{The left panel shows the light curve of an example of each classification type. The right panel shows the phase folded light curves for each classification type. These TICs are all from the TEBC.}
    \label{fig:binary_type}
\end{figure*}
\subsection{Variability Classification}

Following L17, we classified each light curve into the following variability types: starspot modulations, ellipsoidal variations, other periodic variability, and nonperiodic. Figure \ref{fig:binary_type} illustrates examples of the characteristic variability types found in \tess\ data.

\begin{enumerate}
    \item \textbf{Starspot modulations (SP)} appear as quasiperiodic variations in the out-of-transit flux due to dips in the brightness of the star as the spots rotate into and out of view. The shape of the variations changes over time as star spots form and evolve, as well as under the influence of differential rotation. For a star that exhibits differential rotation, star spots at different latitudes imprint different periodicities onto the star’s light curve, resulting in interference patterns that change the shape of the variability over several periods \citep{lurie_tidal_2017}. When phase folded on the orbital period, the flux clearly shows the dips caused by the eclipses but has a shifting appearance of when the flux peaks in the orbit.
    \item \textbf{Ellipsoidal variations (EV)} occur in very close binary systems and appear as two peaks in the light curve halfway between the primary and secondary eclipses. These variations result from the changing apparent cross section of tidally distorted stars as viewed from Earth, which is largest at the quadrature. When phase folded on the orbital period, the flux shows a pattern with broadly winged dips outside of eclipses.
    \item \textbf{Other periodic variability (OT)} are EBs that do not fit in to any of the previous categories, but still show periodic behavior. Some of these may be unidentified heartbeat stars or pulsations in one or both components (such as a $\delta$ Scuti or $\gamma$ Doradus companion). This category also includes non-eclipsing binaries, which are targets that are likely misclassified as EBs (such as fast rotators).
    \item \textbf{Non-periodics (NP)} are EBs that do not exhibit clear out-of-eclipse  variability. Many of these have essentially flat light curves, or variations dominated by noise.
\end{enumerate}

\section{Methods} \label{sec:methods}

\subsection{Classification Procedure} \label{subsec:class}

Since we focus on systems with strong tidal interactions, we filter our sample to targets with orbital periods of $\porb < 10$ d. The dynamical evolution of these short-period systems is strongly impacted by tidal dissipation. These are ideal for \tess observations, as they complete at least two orbits within a single \tess sector, which spans $\sim$ 28 days. Applying this cut left 3684 \tess light curves to classify. Next, we cross-match our filtered sample to the \gaia\ DR3 catalog to obtain the distances, as well as BP, RP, and G magnitude bands to use as features for the classifier (see Table \ref{tab:features}).

\begin{figure*}
    \centering
    \includegraphics[width=0.6\linewidth]{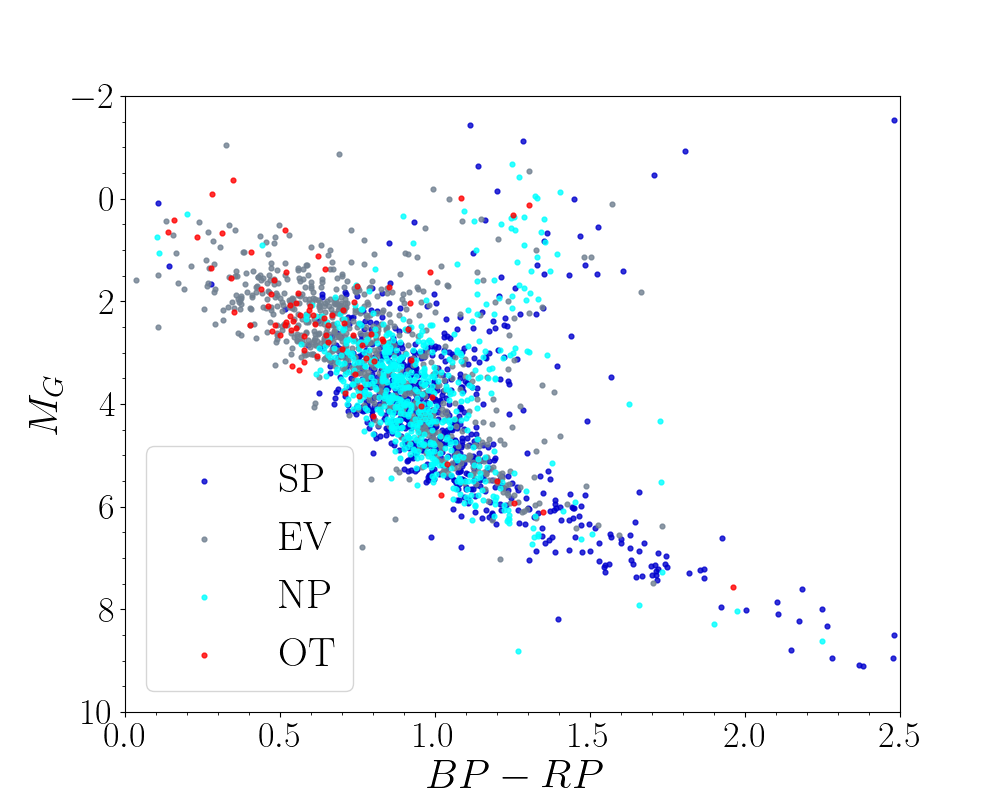}
    \includegraphics[width=0.6\linewidth]{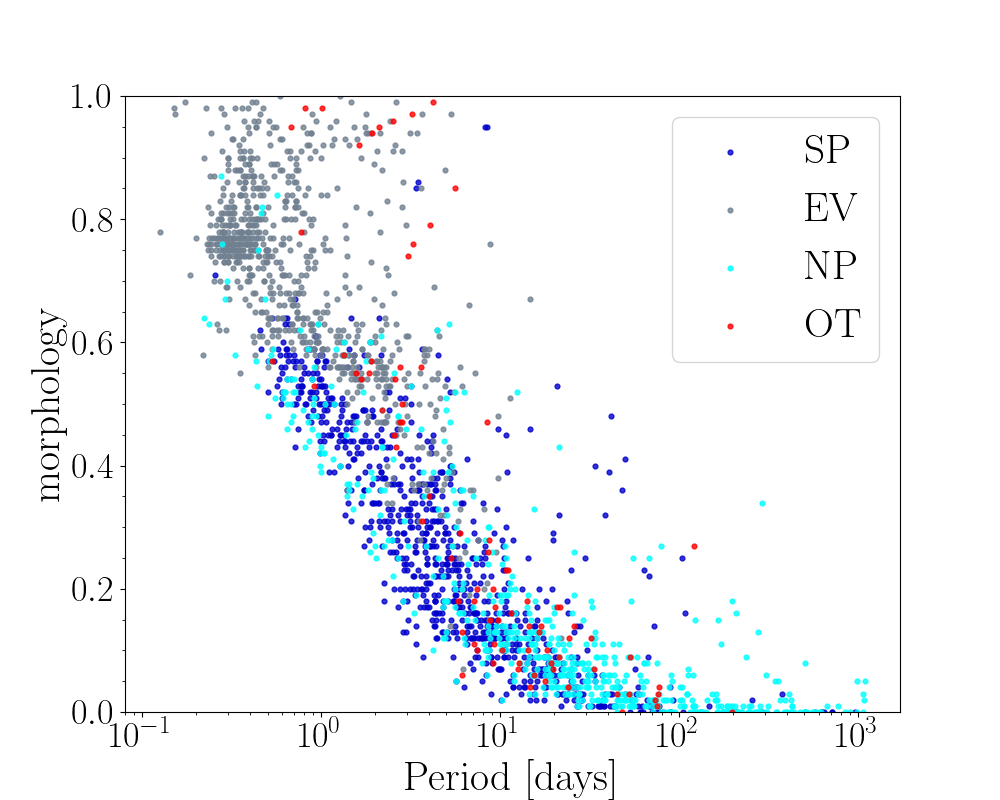}
    \caption{\gaia\ color-magnitude diagram and orbital period vs. morphology of classified \kepler\ EBs from L17. We used this catalog of \kepler\ EBs to train a random forest classifier, which we then applied to classify lightcurves from \tess\ (Section \ref{subsec:class}). Here morphology refers to the metric defined in \citep{matijevic_kepler_2012}, which quantifies how detached a system is ranging between 0 and 1 (where lower value corresponds to more detached).}
    \label{fig:kepler_diagnostics}
\end{figure*}

\begin{longtable*}{rcl} 
\caption{Features used in Random Forest classification. \label{tab:features}}\\
\hline\hline
    amplitude 			&& 		Half the difference between the maximum and minimum magnitude  \\
    $BP-RP$             &&      \gaia color from BP - RP photometry \\
    flux percentile ratio mid 20 && (60 flux percentile - 40 flux percentile) / (95 flux percentile - 5 flux percentile) \\
    flux percentile ratio mid 35 && (67.5 flux percentile - 32.5 flux percentile) / (95 flux percentile - 5 flux percentile) \\
    flux percentile ratio mid 50 && (75 flux percentile - 25 flux percentile) / (95 flux percentile - 5 flux percentile) \\
    flux percentile ratio mid 65 && (82.5 flux percentile - 17.5 flux percentile) / (95 flux percentile - 5 flux percentile) \\
    flux percentile ratio mid 80 && (90 flux percentile - 10 flux percentile) / (95 flux percentile - 5 flux percentile) \\
    $G_{\rm mag}$       &&      \gaia $G$ absolute magnitude \\
    mean absolute deviation &&  Median absolute deviation (from the median) of the observed values \\
    morph               &&      light curve morphology metric \citep{matijevic_kepler_2012} quantifying how detached systems are \\
    percent difference flux percentile && Difference between the 95th and 5th flux percentiles as a percentage of the median value \\
    percent beyond 1 std    && Percentage of values more than 1 standard deviation from the weighted average \\
    period              &&      Orbital period measured from \cite{prsa_tess_2021} \\
    period fast         &&      Period determined from simple sinusoidal fit \\
    std                 && Standard deviation of flux distribution \\
    stetson j           && Robust variance metric \\
    stetson k           && Robust kurtosis statistic \\
    skew                && Skewness of flux distribution \\
    weigted average     && Arithmetic mean of observed values, weighted by measurement errors \\
    RUWE                &&      \gaia Renormalised Unit Weight Error; astrometric goodness-of-fit \\
    \hline\hline
\end{longtable*}

Figure \ref{fig:kepler_diagnostics} shows where the classifications are clustered on a \gaia color-magnitude diagram (left panel) and period vs. morphology plot (right panel). Here we see how the four EB classifications (SP, EV, NP, OT) fall in the two-dimensional parameter space. From the color-magnitude diagram, it is difficult to visually distinguish between the four classifications. More statistical features (e.g. Table \ref{tab:features}), however, can provide additional information to separate different forms of variability and improve the accuracy of classification. On the other hand, high dimensions of features can become difficult to visualize and cumbersome to interpret and classify, motivating the use of an automated machine learning approach.

Machine learning has been widely successful in automating the classification of different types of stellar variability (and other astrophysical variability) that is observable in photometric time series \citep{richards_machine-learned_2011,Kim_2016,jayasinghe_asas-sn_2018}. Given a well-classified sample of light curves, studies such as \cite{jayasinghe_asas-sn_2018} have shown that light curve feature extraction and random forest (RF) classification are effective at distinguishing eclipsing binaries from other forms of astrophysical variability,  as well as differentiating detached binaries from contact binaries. For this reason, we chose the RF method among the many machine learning classifiers. A detailed comparison of different classification methods on stellar variability is beyond the scope of this paper \citep[see][for a comprehensive review]{yu_survey_2021}.

Our classification of the \tess EB sample consisted of two steps: an initial classification using machine learning, followed by human vetting. Training was performed by extracting features from \kepler\ lightcurve data, and then applied to features extracted from \tess light curve data. This involved training a RF classifier on \kepler\ lightcurves of EBs with known labels from L17, then applying the RF classifier to previously unclassified \tess\ EBs. Although the machine classifier provided mostly accurate classifications, it was limited by potential differences of extracting features from two different surveys (e.g., different cadence, noise, and systematics). To more thoroughly inspect for catalog purity, we followed up with a vetting phase in which we manually inspected and validated each light curve in the sample. 

For both \kepler and \tess EB samples, we compute statistical features of each light curve and cross-match the samples with \gaia DR3 \citep{gaia_dr3}, resulting in a sample of 2182 \kepler EBs with known classifications, and 3684 \tess EBs with unknown classifications.
To compute the statistical features from the light curves we used the Python package \code{cesium} \citep{cesium}. A description of the 20 features included in our classifier is summarized in Table \ref{tab:features}. 

\begin{figure*}
    \centering
    \includegraphics[width=0.5\linewidth]{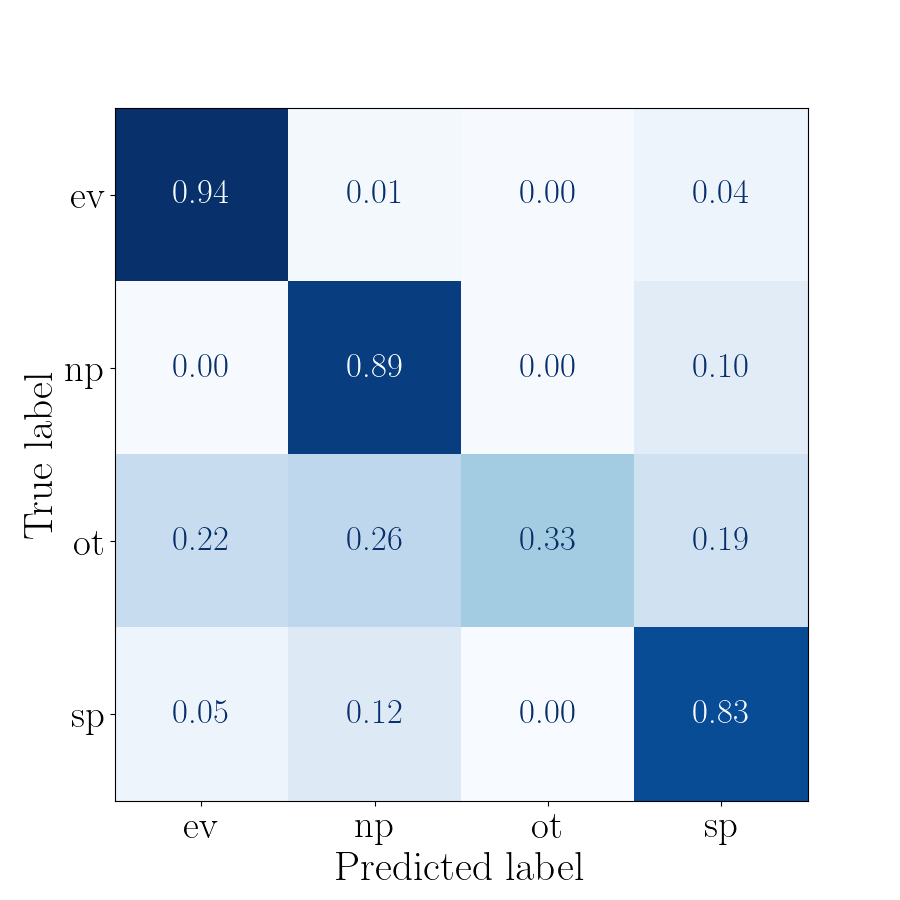}
    \caption{Confusion matrix evaluating the accuracy of our random forest classifier trained on labels of eclipsing binaries classified by L17. Each light curve is classified into 4 categories: ellipsoidal variability (ev), non-periodic (np), other variability (ot), and starspot variability (sp). The y-axis shows the true label (as classified by L17) and the x-axis shows the label predicted by the random forest classifier. Entries along the diagonal of the matrix represent the percentage of sources that were accurately labeled out of the total number of true labels in each category.} 
    \label{fig:rf_matrix}
\end{figure*}

\begin{figure}
    \centering
    \includegraphics[width=0.5\linewidth,origin=c]{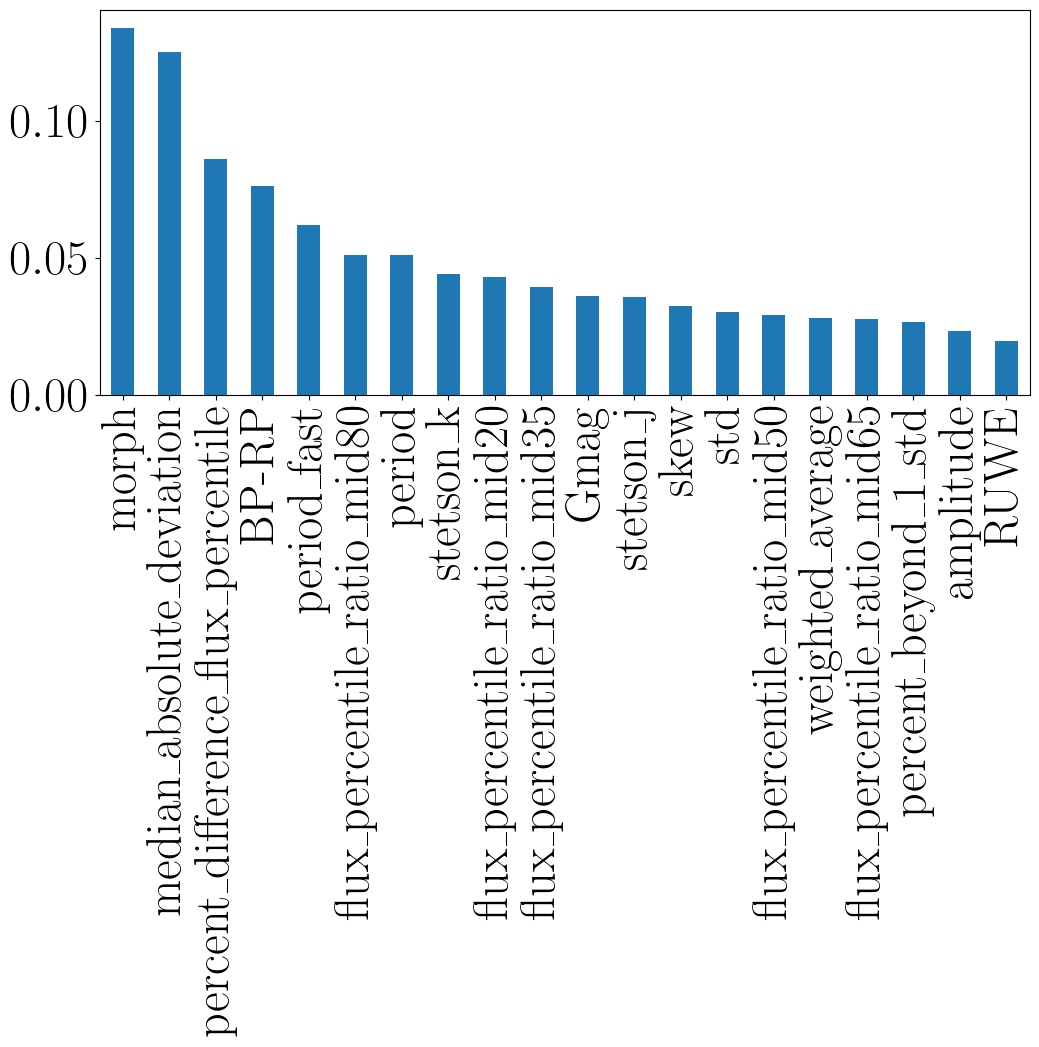}
    \caption{Random forest feature importance with descriptions of each feature given in Table \ref{tab:features}.}
    \label{fig:rf_importance}
\end{figure}

For the training procedure, we randomly divided the sample of 2182 \kepler EBs into \emph{training sample} with 65\% of the sources ($N_{\rm train}=1418$), and a \emph{test sample} with the remaining 35\% of the sources ($N_{\rm test}=764$), which we found reduced the risk of overfitting. To choose the hyperparameters of the random forest classifier and avoid overfitting the model, we subdivided the training sample of 1418 sources and performed a 5-fold cross-validation test \citep{yates_cross_2022}. From the cross-validation test, we set the number of random forest estimators to $N_{\rm estimators}=128$, which when applied to the test sample of 764 sources, achieved a test-sample accuracy of 85\%. The confusion matrix plotted in Figure \ref{fig:rf_matrix} shows the accuracy breakdown for each of the four classes of EBs. 
Figure \ref{fig:rf_importance} shows the relative importance of each feature described in Table \ref{tab:features}. 

\begin{figure}
    \centering
    \includegraphics[width=0.5\linewidth]{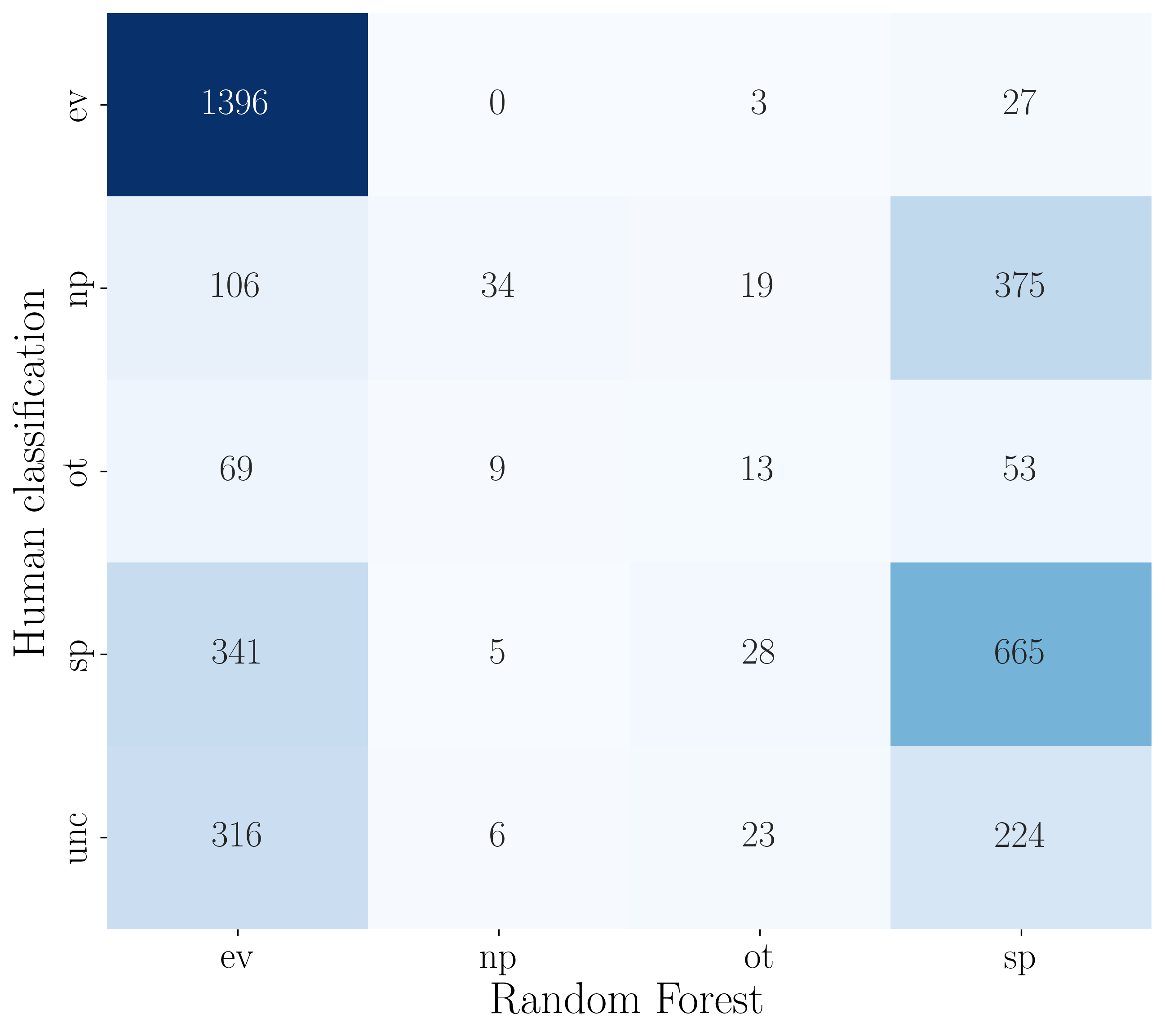}
    \caption{Classes predicted by the random forest classifier compared to the human verified classifications.}
    \label{fig:verified_class}
\end{figure}

We further thoroughly inspected the entire sample by visually vetting each light curve by at least two people. Figure \ref{fig:verified_class} shows the classifications after human vetting. This visual verification procedure left a sample of 1039 candidate SP EBs. 
\subsection{Period Measurement} \label{subsec:period}
To measure rotational periods, we use the Lomb-Scargle (LS) periodogram, autocorrelation function (ACF), and phase dispersion minimization (PDM) function. The LS periodogram calculates the normal Fourier power spectrum and uses the least squares method to determine the best measure of the period \citep{lomb_least-squares_1976}. The LS method is advantageous in that it is efficient to compute for dense time-series and applicable to unevenly sampled data. However, LS also assumes that the time-series signal can be decomposed as a summation of sinusoidal signals, which does not always perform well at representing the morphology of the eclipsing or star spot variability. For this reason, we also apply two other period-finding algorithms.

The autocorrelation function (ACF) for signal processing calculates the autocorrelation coefficient at time lag $k$ for time series $x_i$, written in Equation (\ref{eqn:ACF}) \citep{ACFBook} as
\begin{equation} \label{eqn:ACF}
    ACF_k = \frac{\sum_{i=1}^{N-k} (x_i-\bar{x})(x_{i+k}-\bar{x}) }{\sum_{i=1}^{N} (x_i-\bar{x})^2}.
\end{equation}
Where the maximum peak in ACF values should correspond to the rotational period. ACF is advantageous when the amplitude and phase of photometric modulation significantly evolve in observations because the period remains detectable. 
Peaks in the ACF can also occur at integer factors of the true period (often emerging at double or half the actual period). As a result, we opted for visual selection rather than solely relying on the assumption that the first peak always represents the rotational period.

The phase dispersion minimization (PDM) approach \citep{stellingwerf_period_1978} is a nonparametric method in which the period is estimated by minimizing the scatter for the phase-folded light curve. Our implementation for PDM works as follows. First, we detrend long-term (non-rotational) variability signals using a low-order polynomial fit \citep{detrend_welsh_1999}. The light curve is then phase-folded on a trial period $P_0$, where $\phi(P_0)$ denotes the phased flux. Next the phase-folded light curve is smoothed using a rolling median function centered at the $i^{th}$ timestep with a width $w$, which is computed as $\phi_{\rm median}(t_i, P_0) = {\rm median}(\phi(t_{i - w}, P_0), \dots, \phi(t_{i+w}, P_0))$. Next we compute the scatter between the flux and the median curve, and optimize for the period, which minimizes the weighted scatter:
\begin{equation} \label{eqn:pdm}
    P_{\rm pdm} = \operatorname*{argmin}_{P_0} \sum_{i=0}^{N} \left( \frac{\phi(t_i, P_0) - \phi_{\rm median}(t_i, P_0)}{\sigma_i} \right)^2.
\end{equation}
As the PDM method is computationally expensive to perform for densely sampled time series, we do not compute the phase dispersion for a wide range of trial periods. Instead, we initialize the trial period using the visually inspected LS or ACF period and use PDM to improve the precision of the period. Optimization of the weighted scatter, Equation (\ref{eqn:pdm}), is performed numerically using \code{scipy.optimize}. While computationally efficient, this initialization biases the results toward the chosen range, potentially converging on a local rather than the true global minimum in scatter. In most cases this improves the precision of the period measurements over the LS and ACF methods, but does not account for possible systematic biases, such as multiple starspots, which can make it challenging to identify the true rotational period.


\subsection{TESS Light Curves}
\begin{figure}
    \centering
    \includegraphics[scale=0.22]{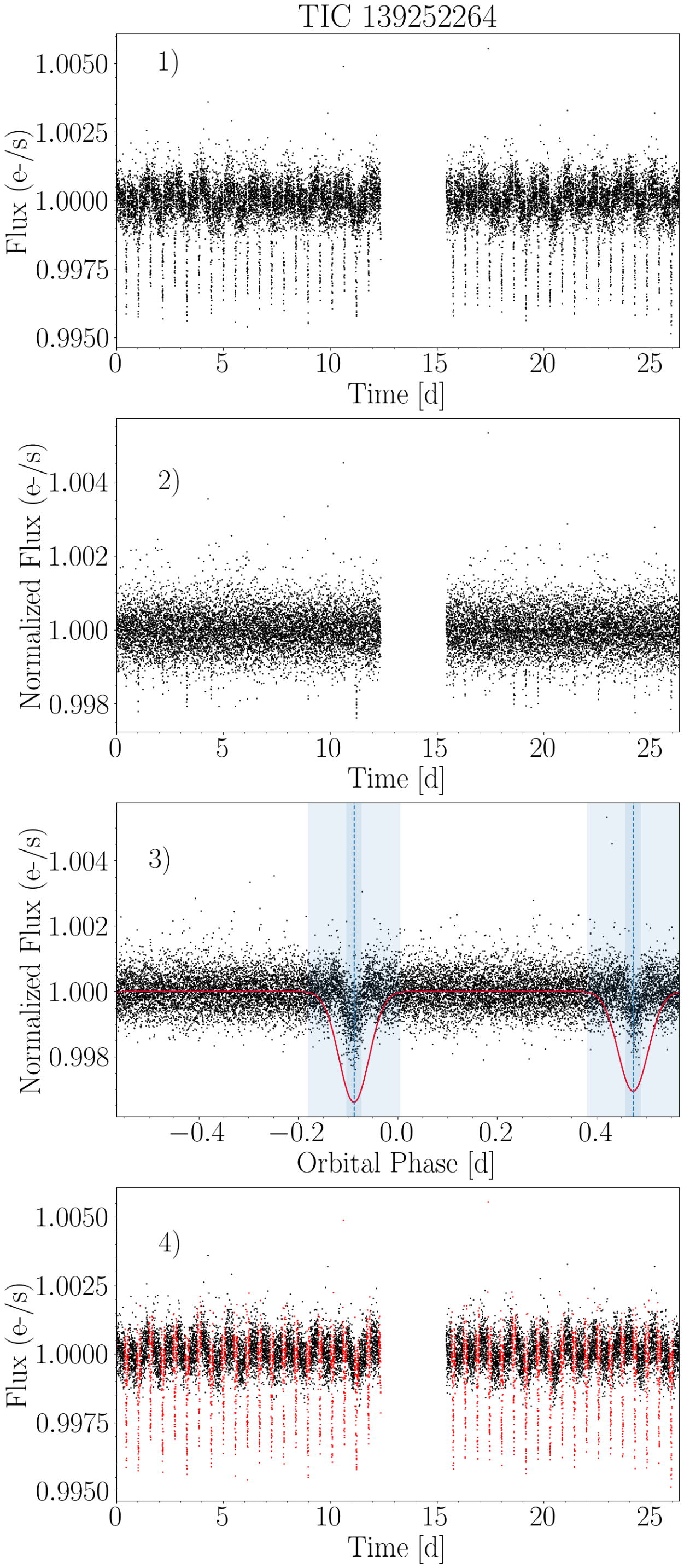}
    \caption{Example of a light curve with verified eclipses and masked-out data points. The top panel is the original light curve. The second is the normalized light curve. The third shows the phase-folded light curve at the optimized orbital period ($P_{\text{orb}} = 1.132$ days), with eclipses modeled as Gaussian functions ($A_1 = 0.003$, $t_1 = 0.088$, $\sigma_1 = 0.03$; $A_2 = 0.003$, $t_2 = 0.474$, $\sigma_2 = 0.03$) and subtracted to generate the mask.} The fourth shows the original light curve with the mask applied to the eclipses (red dots). Panels are numbered in the top-left corner, for cross-reference in the text.
    \label{fig:4panel}
\end{figure}
We analyze light curves in sectors 1-26 of the \tess\ survey from the \cite{prsa_tess_2021} catalog, and used \lk\ to download all sectors of the data from the \tess\ SPOC pipeline (Figure \ref{fig:4panel}, Panel 1). To extract eclipse variation, we detrend the rotational variability by normalizing the light curve with the \lk flatten function, which splits the light curve into segments and individually applies the \code{SciPy} savgol$\_$filter (Figure \ref{fig:4panel}, Panel 2). Once normalized, we then phase-folded the light curves on the orbital periods listed in \cite{prsa_tess_2021}, using the \lk\ fold method (Figure \ref{fig:4panel}, Panel 3). With these light curves, we were able to map models of each binary star's eclipse by fitting Gaussian models to dips in the flux from the eclipses of both stars in the binary system \citep{mowlavi_gaia_2017}. The phase-folded flux value for the eclipses was calculated using Equation (\ref{gaussian}). Where $i$ indicates the primary ($i=1$) or secondary ($i=2$) eclipse, $A$ is the amplitude of the dip in phase-folded flux, $t$ are the time values in the phase-folded light curve, $t_{\rm min}$ is the time of the dip's minimum, and $d$ is the duration of the dip in days:

\begin{equation}
    y = A_i  \cdot  \exp\left[-(t - t_{\rm min, i})^2/(2d_i^2) \right]
    \label{gaussian}.
\end{equation}

We used the minimum of the phase-folded light curve flux values to obtain the primary eclipse amplitude ($A_1$) and the time at the minimum ($t_{\rm min,1}$). This technique was repeated to identify the second minimum, resulting in the secondary eclipse's amplitude ($A_2$) and time at the secondary minimum ($t_{\rm min,2}$). In this initial optimizer, the duration ($d_1$ and $d_2$) for both eclipses was set to the same fixed value (typically 0.03 days) and $t_{\rm min,2}$ was assumed to be within 0.25 days in phase of $t_{\rm min,1}$. These estimations were adjusted manually, if required, after an inspection of an over-plot of the eclipse models on the phase-folded light curve. We ran these parameters through a second optimizer that minimized $\chi^2$ through the least squares method in the bounds of the phase-folded light curve. This yielded our final parameters for the eclipse models, as well as an optimized orbital period (Figure \ref{fig:4panel}, Panel 3).

With the optimized parameters, we then mask the overlap of the eclipse models on the original data, which removes the flux measurements during eclipses that creates high scatter in phase-folded light curves that are not folded on the orbital period. To remove the eclipse signals from the light curves, we used a duration mask of $\pm3$ standard deviations from the mean of the Gaussian fit. This mask was made using the eclipse mask of \code{BoxLeastSquares} using the eclipse times ($t_{\rm min,1}$ and $t_{\rm min,2}$), the best-fit orbital period (obtained from LS, ACF and PDM in Section \ref{subsec:period}), and the eclipse duration ($d_1$ and $d_2$). Two masks (one around each eclipse at $t_{\rm min,1} \pm 3 d_1$ and $t_{\rm min,2} \pm 3 d_2$) were made and applied to each of the data arrays for time, flux, and error. These masks are the red scatter values in Figure \ref{fig:4panel}, Panel 4.

\subsection{Measurement Procedure Rotational Periods}

\begin{figure}
    \centering
    \includegraphics[scale=0.4]{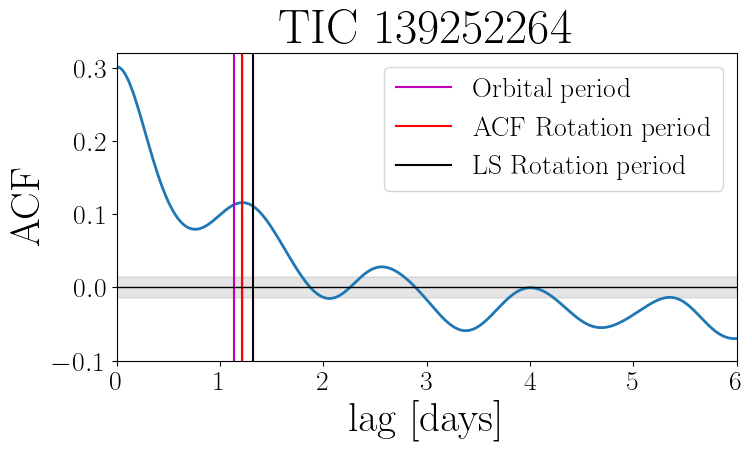}
    \includegraphics[scale=0.4]{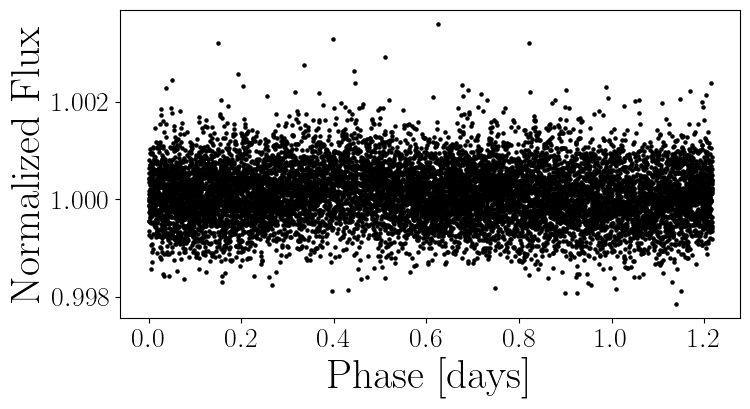}
    \includegraphics[scale=0.4]{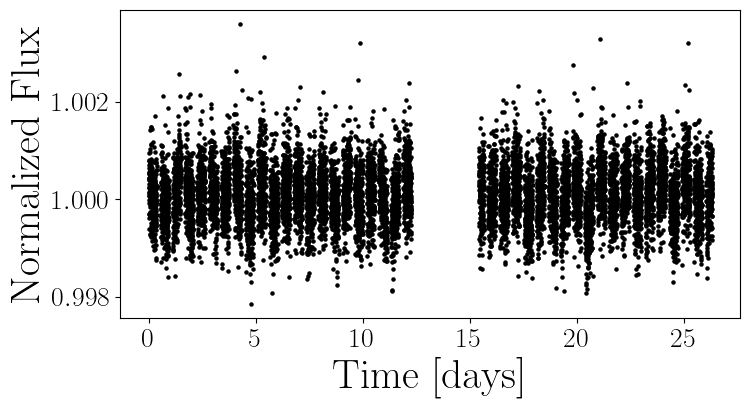}
    \caption{Example of a light curve with an applied mask on the orbital period and eclipses, used with the autocorrelation function (ACF) to identify the rotational period. The first plot is the ACF periodogram, with the orbital period, LS rotational period, and ACF rotational period marked. The second is the light curve phase folded on the ACF period. The third is the original light curve with eclipses masked out.}
    \label{fig:ACF}
\end{figure}

\begin{figure}
    \centering
    \includegraphics[scale=0.4]{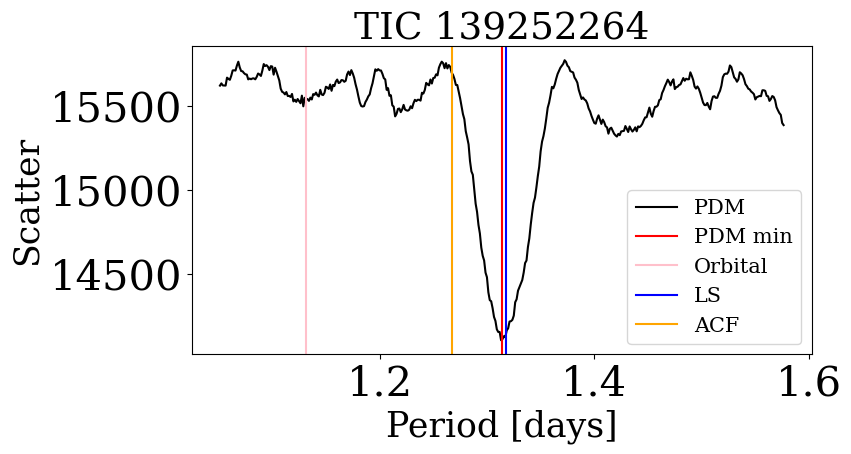}
    \includegraphics[scale=0.4]{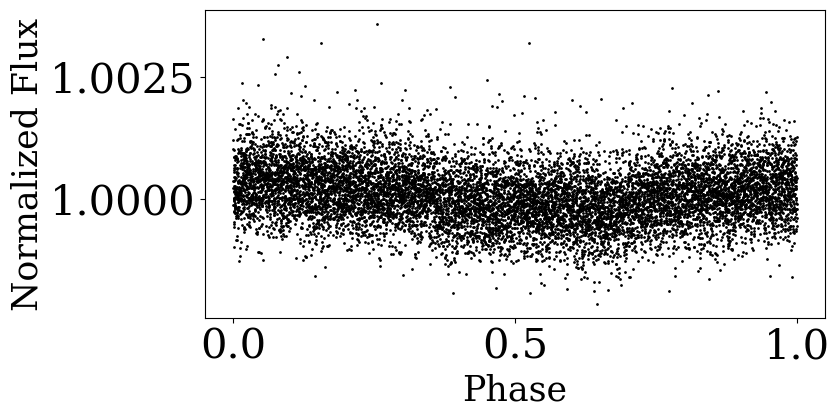}
    \caption{Example light curve (TIC ID 139252264) with an applied mask on the orbital period and eclipses, used in conjunction with phase dispersion minimization (PDM) to identify the rotational period. The first plot shows a PDM periodogram comparison, with the orbital period, LS rotational period, and ACF rotational period marked. The second displays the light curve phase-folded on the PDM identified rotational period.}
    \label{fig:PDM}
\end{figure}

After measuring the orbital periods using the model in \ref{subsec:period}, we measure the rotation periods. First, we mask out the eclipses from the light curve using the eclipse timings and durations from our transit model fit \ref{gaussian}. Next, we apply both Lomb-Scargle and ACF to the masked light curve to measure the rotation period as described in Section \ref{subsec:period}. In cases where the highest peak in the ACF does not look like the rotational period, we manually specify the local maximum to use as the output from the ACF (Figure \ref{fig:ACF}, Panel 1). If neither the LS nor ACF measurements appear correct, we also apply PDM. In the PDM, we use the results of the rotational period of the LS and ACF methods as an initial guess (Figure \ref{fig:PDM}, Panel 1).If neither LS nor ACF provides a suitable initial guess, we estimate the period manually by visually inspecting the original light curve.

To visualize whether any of these values represented the correct rotational period, we plot phase-folded light curves (Figure \ref{fig:4panel}, Panel 4; Figure \ref{fig:ACF} Panel 2; Figure \ref{fig:PDM}, Panel 2) and visually inspected these plots for each method. The rotational period value that produces the most periodic-looking single phase among the methods is saved as the rotational period for the TIC ID. There were outlier cases where the phase-folded light curve did not appear periodic or suggested the presence of an n-body system. We discarded these outliers from our sample catalog because we were unable to identify a rotational period using our described methods (Section \ref{subsec:period}).

\subsection{Bayesian model comparison}\label{bayesian}

Next, we analyze the $P_{\rm orb} / P_{\rm rot}$ distribution of the sample and use Bayesian model comparison to quantify the significance of the second subsynchronous peak discovered in L17. To perform this comparison, we calculated whether a double-peaked model was a significantly better fit than a single-peaked model. Given that we do not have a theoretical model for the spin-orbit distributions (and as discussed in the results, the empirical distributions are roughly symmetric), we chose to fit each distribution using a combination of either Gaussian or Lorentzian functions. In particular, we evaluated six models: single Gaussian, single skewed Gaussian, double Gaussian, single Lorentzian, single skewed Lorentzian, and double Lorentzian. For the models, we use the \code{scipy} norm, skewnorm, cauchy and skewcauchy functions with the cdf method for a single Gaussian, single skewed Gaussian, single Lorentzian, and single skewed Lorentzian, respectively. The double peak models consisted of two of the same functions, the second of which had an amplitude parameter, which were summed and normalized.

Rather than using a histogram, which may miss the intricacies of the data, we chose Kernel Density Estimation (KDE) to estimate a Probability Density Function (PDF). Using \code{GridSearchCV} from the package \code{sklearn} and 20-fold cross-validation to ensure low coefficient variability, we calculated the ideal bin width for the KDE yielding a smoothed PDF. Following that, we normalize the PDF by dividing it by its sum of values.

The challenge with using a KDE is that its effectiveness depends heavily on the chosen bin width. Oversmoothing may occur with a large bin width, while a small width may result in undersmoothing. Despite KDE's ability to provide useful PDF visualizations, we decided to convert the sample into an estimated Cumulative Density Function (CDF) because it is derived entirely from the data sample, which eliminates worries about artificial feature generation or manipulation that might impair model fitting. The CDF function was calculated by dividing the cumulative total of unique spin-orbit ratios by their sum. To test whether it is plausible that the \kepler and \tess samples originate from the same underlying distribution, we perform a two-sample Cramér-von Mises test \citep{cramer_von_mises_1962} with the null hypothesis that the two populations are identical. Applying this test to the two CDF functions, we compute an output p-value of 0.99, indicating that we cannot reject the null hypothesis.

There are a couple paths forward, and with this p-value we decided to handle this with this next methodology.

This test suggests that the samples originate from the same underlying distribution. To evaluate the statistical significance of the subpopulation, we modeled the \tess, \kepler, and combined data samples. 

\begin{table}[h]
\begin{center}
\begin{tabular}{|c|c|c|}
    \hline
    Parameter & Minimum & Maximum \\
    \hline 
    $\mu_1$ & 0.9 & 1.1\\
    $\sigma_1$ & 0.01 & 0.5\\
    $\gamma_1$ & 0.001 & 0.5\\
    $\mu_2$ & 0.7 & 0.9\\
    $\sigma_2$ & 0.01 & 0.05\\
    $\gamma_2$ & 0.01 & 0.05\\
    $a$ & 0.1 & 1\\
    $s_g$ & -4 & 4\\
    $s_l$ & -1 & 1\\
    \hline
\end{tabular}
\caption{The ranges for Gaussian and Lorentzian parameters, listed with the minimum and maximum for the ranges. The parameters are $\mu_1$ (primary mean), $\sigma_1$ (primary standard deviation), $\gamma_1$ (half width half maximum), $\mu_2$ (secondary mean), $\sigma_2$ (secondary standard deviation), $\gamma_2$ (half width half maximum), a (secondary amplitude), $s_g$ (Gaussian skew), $s_l$ (Lorentzian skew). The primary amplitude is always 1, so it is not included as a parameter.}
\label{tab:parameter_bounds}
\end{center}
\end{table}

To identify optimal model fits, we implement a Bayesian sampling approach using the \code{dynesty} \code{NestedSampler}. We sample the combined data using the natural log-likelihood of single-, skewed and un-skewed, and double-peaked Gaussian and Lorentz CDFs, with each sampler assuming a uniform prior (given the ranges in Table \ref{tab:parameter_bounds}). Since the measurement error for the orbital and rotational periods is unknown, we apply a uniform error of 0.025 to all spin-orbit ratios in the likelihood functions, which is equivalent to the standard deviation of a single Gaussian KDE on the \tess data.

One limitation of the period measurement methods we employed (Lomb-Scargle, autocorrelation function, and phase dispersion minimization, Section \ref{subsec:period}) is that it is difficult to accurately determine the uncertainty of the period measurement. We estimate a fiducial error of $\sigma = 0.025$, based on the standard deviation of sources around the 1:1 spin orbit ratio (assuming that most sources around 1:1 are synchronized and the dispersion around 1:1 is mostly due to measurement error). However, since this value for $\sigma$ is only a rough estimate, we later test the significance of the Bayes factor as a function of error (mentioned at the end of this section) by varying the error values between the range of $\sigma = 0.01 - 0.5$. A rigorous analysis would incorporate period-specific errors, but deriving these would require computationally expensive individual modeling of each light curve. For the statistical analysis presented here, we adopt a uniform error as a conservative and computationally tractable approximation, capturing the expected range of uncertainties without presupposing an unconstrained error distribution.

For our six different models, $Y$ (single Gaussian, single skewed Gaussian double Gaussian, single Lorentzian, single skewed Lorentzian, and double Lorentzian), we calculated the natural log-likelihood of the model fit as: 
\begin{equation}\label{eq:ln-likelihood}
    \ln\mathcal{L}(Y | \theta) = -\frac{1}{2} \sum_{i=1}^N \left[\frac{Y(x_i) - C(x_i)}{\sigma} \right]^2 - 2 N \ln\sigma,
\end{equation}
where $x_i = P_{\rm orb} / P_{\rm rot}$ is the spin-orbit ratio, $C(x_i)$ is the CDF of the data as a function of spin-orbit ratio, $Y(x_i)$ is the model fit, and $\sigma$ is the error for the observations $x_i$.

The bimodal posteriors include an amplitude parameter to scale to the secondary peak observed in the subsample. The ranges for the parameters are listed in Table \ref{tab:parameter_bounds}. It should be noted that for the single Lorentz we changed the range of $\gamma_1$ to have a maximum of 0.02, because the sampler made $\gamma_1$ much too large when sampling. The bounds for $\mu_1$ and $\mu_2$ are the only observationally constrained boundaries, the remaining are arbitrary. We see from Figure \ref{fig:conf_KvT_1} that the primary mean should be a spin-orbit ratio around 1, and the secondary mean should be a spin-orbit ratio around 0.8. Since we don't know the exact value, we give a buffer of 0.1 in both the additive and subtractive directions.

To assess which posterior best aligns with the combined data, we computed the Bayes factors, representing the marginal likelihood ratio of two different models \citep{doi:10.1080/01621459.1995.10476572}. The marginal likelihood (or evidence) $Z$ is computed by integrating the model likelihood over the prior:
\begin{equation} \label{eq:evidence}
    Z = P(D | M) = \int P(D | \theta, M) \, P(\theta | M) \, d\theta,
\end{equation}
where $D$ is the data and $M$ is a given model with parameters $\theta$, which are different for each of the four models (Table \ref{tab:parameter_bounds}): $\theta_{sg} = \{\mu, \sigma \}$ for single Gaussian; $\theta_{dg} = \{\mu_1, \sigma_1, \mu_2, \sigma_2, a \}$ for double Gaussian; $\theta_{sl} = \{\mu, \gamma \}$ for single Lorentzian; and $\theta_{dl} = \{\mu_1, \gamma_1, \mu_2, \gamma_2, a \}$ for double Lorentzian. Subsequently, the natural log-marginal likelihood was utilized in Equation (\ref{eq:BF}) to compute the natural log Bayes factor,
\begin{equation}\label{eq:BF}
\begin{split}
    \mathcal{B} &= \frac{Z_1}{Z_2},\\
    \ln\mathcal{B} &= \ln Z_1  - \ln Z_2.
\end{split}
\end{equation}
The results of our model comparison test are discussed in Section \ref{sec:bayes_factor_results}. \\

\section{Results} \label{sec:results}

In this section, we give a brief overview of our rotation period catalog and the subsample of EBs classified with starspot rotation. We then use this subsample to inspect the synchronization trend of EBs classified with starspot rotation.

\subsection{Rotation Period Catalog}

Table \ref{tab:catalog} lists the column titles and their descriptions for our rotational period catalog. The full catalog is available in the online supplement. We provide the eclipse and duration parameters along with the orbital period needed to mask out eclipses from the \tess light curve. We also include all measurements of the rotational period across methods and separately list the rotational period we use in our analysis (rotation period inspected).
\clearpage

\begin{longtable*}{rcp{5in}} 
\caption{Columns of Catalog \label{tab:catalog}}\\
\hline
    Name & Units & Description\\
\hline
    TIC & - &\tess\ object ID\\
    
    primary\_eclipse\_amplitude& - &Amplitude of primary eclipse\\
    
    primary\_eclipse\_time& days &Phase of primary eclipse\\
    
    primary\_eclipse\_duration& days &Duration of primary eclipse\\
    
    secondary\_eclipse\_amplitude& - &Amplitude of secondary eclipse\\
    
    secondary\_eclipse\_time& days &Phase of secondary eclipse\\
    
    secondary\_eclipse\_duration& days &Duration of secondary eclipse\\
    
    orbital\_period& days &Resultant orbital period from initial optimizers\\
    
    LS\_rot\_period& days &Rotational period found using the Lomb-Scargle method\\
    
    flag\_note& - &Issues encountered with the TIC\\
    
    rot\_period\_inspected& days &Rotational period from visual inspection (adopted as the true period)\\
    
    ACF\_rot\_period& days &Rotational period found using the autocorrelation function\\

    PDM\_rot\_period& days &Rotational period found using phase dispersion minimization\\
    
    synch& - & orbital period / rotation period inspected\\
    
    alias\_period& days &Period identified in at least one method as the rotational period, but was within a half or double increment of the inspected period\\
    
    eccentricity& - & $\pi/2(|(\text{primary transit time} - \text{secondary transit time)/orbital period}|-0.5)$\\
    
    nsectors& - &Number of sectors used when calculating LS and ACF rotational periods\\

\hline
\end{longtable*}

\subsection{Starspot Rotation subsample}\label{subsec:subsample}

To ensure confident calculation of a system's rotational period, we created a subsample of EBs classified as starspot rotation, comprising 584 TIC IDs. To fit the orbital period, we used all sectors of data with the TIC IDs. We then used the methods described in Section \ref{sec:methods} to obtain a fitted value for the orbital period, LS rotational period, and ACF rotational period. We initially used a single sector of data, and then used full-sector data for EBs with orbital periods $>$6 days. We opted not to employ the PDM rotational method for the full-sector data due to the extensive calculation time required. Instead, we used data solely from the first sector for PDF rotational periods. The inspected rotational period compares the full sector ACF and LS periods with the single sector PDM period. The period with the lowest scatter and most obvious rotation period by visual inspection was saved as the inspected period.

Our visual inspection consisted of comparing the phase-folded light curves for each method of rotation period measurement. If the scatter values across methods were similar (showing no significant differences in the first 3-5 orders of magnitude) we made a judgment based on the most apparent rotational period. For example, when one phase-folded light curve appeared jagged, flat, or noisy, while another displayed a relatively smooth curve, we selected the rotational period associated with the smoother curve.

To analyze the degree of synchronization for a given EB, we calculate the spin-orbit ratio of $P_{\rm orb}/P_{\rm rot}$. Synchronization occurs at $P_{\rm orb}/P_{\rm rot}=1$; where $P_{\rm orb}/P_{\rm rot}<1$ is subsynchronous, and $P_{\rm orb}/P_{\rm rot}>1$ is supersynchronous. For our subsample, 88\% of the EBs have period ratios in the range of 0.75 to 1.25. To ensure accurate results, we recomputed the rotational and orbital periods using full sector data for TICs that fall within the subsynchronous range $0.82 \lesssim P_{\rm orb}/P_{\rm rot} \lesssim 0.92$. This ratio is used throughout our figures to find significant correlations with synchronization.

\subsection{Comparing rotational period measurements from different methods}

We evaluated the overall precision of each method using two metrics: the root mean squared error (RMSE) and 10\% accuracy of the methods. The RMSE was calculated by taking the root mean square of the difference of the inspected period and a statistical method's period. The root mean squared error for LS was 4.3 days, ACF was 1.5 days, and PDM was 0.2 days. This metric is strongly skewed by outliers, which are most prevalent in the LS method.

\begin{figure}
    \centering
    \includegraphics[scale=0.6]{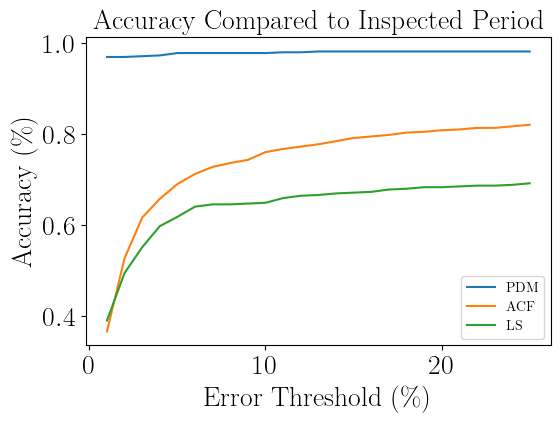}
    \caption{Plot of each method's level of accuracy against an error threshold range of 1-25\%. The error is the root mean square of the difference of the method rotational period to the inspected rotational period.}
    \label{fig:accuracy}
\end{figure}

Meanwhile, 10\% accuracy is less affected by outliers, and suggests which method results in higher accuracy. Examining 10\% accuracy ensures that we are considering the correct period and not an alias that can result from orbital effects. When assessing how often the methods were within 10\% of the inspected period, LS had 64\%, ACF had 76\%, and PDM had 98\% of values with 10\% accuracy. Figure \ref{fig:accuracy} shows accuracy trends of each method for a given error threshold, with the error ranging 1-25\%. By inspecting these three methods, we found that the PDM period accounted for 96\% of the inspected periods.

\begin{figure*}
    \centering
    \includegraphics[scale=0.59]{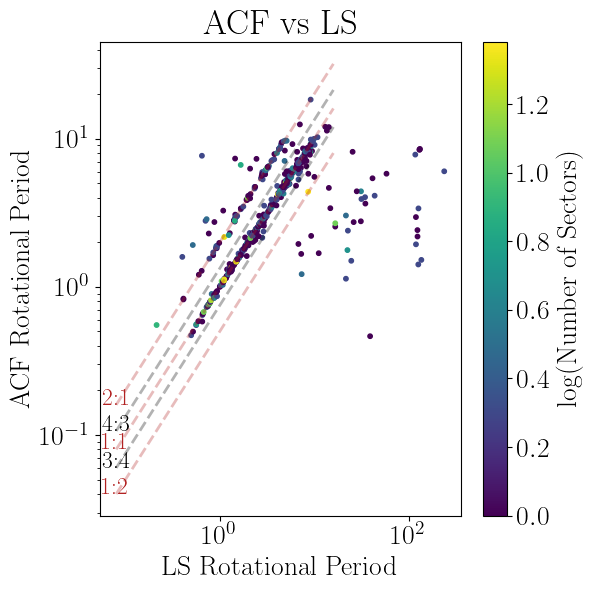}
    \includegraphics[scale=0.59]{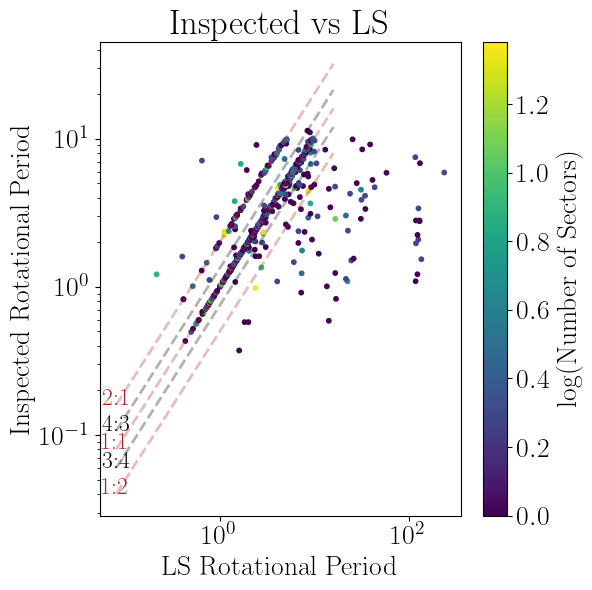}
    \includegraphics[scale=0.59]{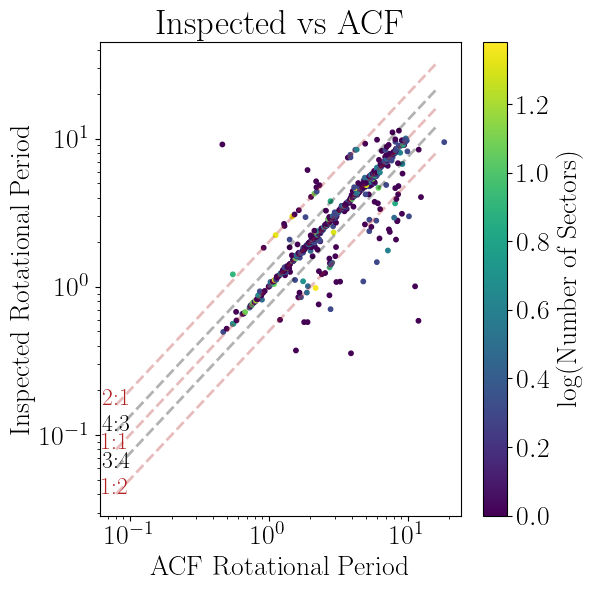}
    \caption{Comparison of rotational period measurements from different techniques, with dashed lines indicating common ratios.}
    \label{fig:tech_result_comp}
\end{figure*}

We also compare our different methods of measuring the rotational period to see if there are any trends that bias our measurements in a method. Figure \ref{fig:tech_result_comp} compares our methods to one another in log-log space, with given ratios marked with dashed lines. Out of all the techniques, the LS method is the least accurate in that it overestimates the rotational period, as well as often doubling the period when compared to both the ACF and inspected period. This is because Lomb-Scargle is moved out of phase as star spots evolve, making ACF overall more reliable for star spot modulated lightcurves \citep{gordon_stellar_2021}. When comparing the ACF and inspected periods, note that the inspected period largely corresponds to the PDM period, which was measured with only a single data sector. Data points with more than one sector could be more accurately measured using the ACF method. The ACF method appears to largely align with the inspected period, although there are some trends indicating doubling or halving of the rotational period.

\begin{figure*}
    \centering
    \includegraphics[scale=0.6]{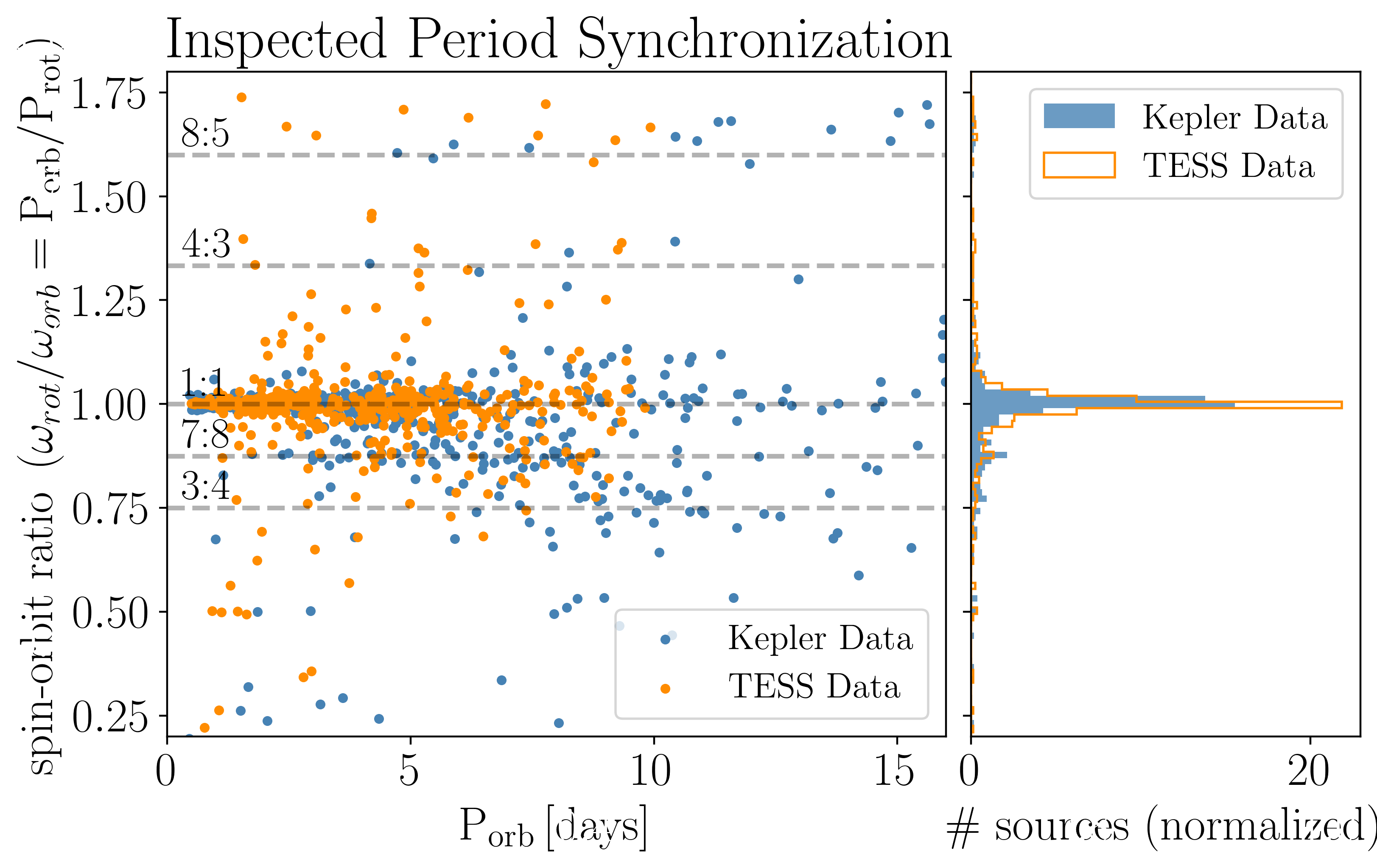}
    \caption{Full \tess sample inspected period, plotted alongside the L17 \kepler data.}
    \label{fig:conf_KvT_1}
\end{figure*}

\begin{figure*}[ht]
    \centering
    \subfigure{\includegraphics[scale=0.42]{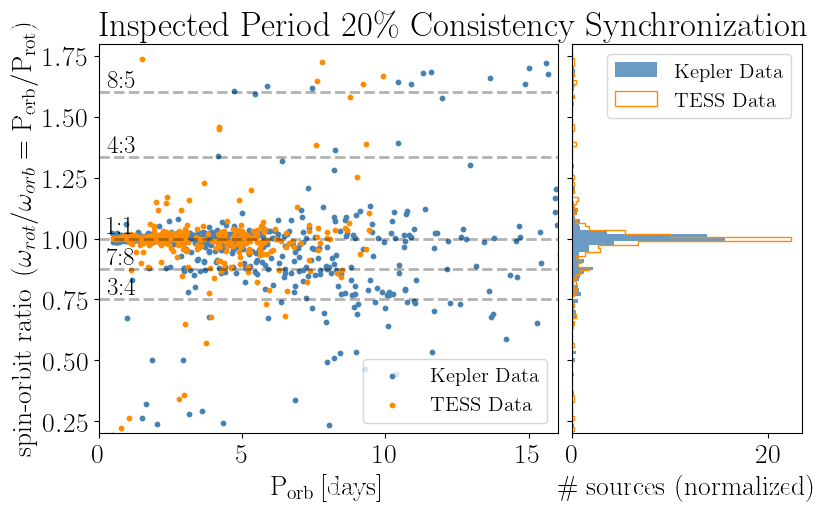}} \subfigure{\includegraphics[scale=0.42]{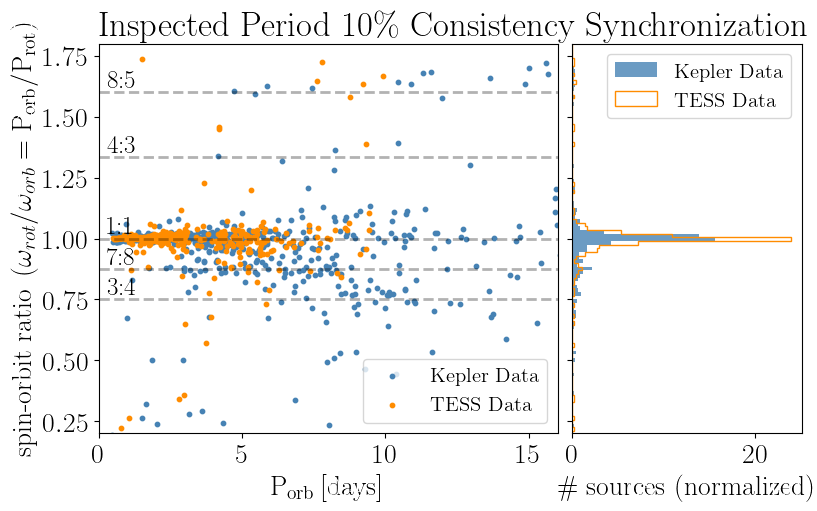}}
    
    \caption{Left panel: \tess subsample of values with 20\% cross-method confidence of the inspected period, plotted alongside the L17 \kepler data. Right panel: Same as left figure, but with 10\% cross-method confidence.}
    \label{fig:conf_KvT_2}
\end{figure*}

To better understand the results, we filter our sample for the most consistent results across the methods. Figure \ref{fig:conf_KvT_1} plots the period ratio of the inspected period alongside the L17 data. See Figure \ref{fig:conf_KvT_2} for a look at the sub-sample at different confidence levels, where we see the subsynchronous secondary peak is only prominent in the consistency range of 20\%. For our full sample in Figure \ref{fig:conf_KvT_1}, the percentage of the subpopulation to the total population is 6\% and emerges as a weak subsynchronous peak in the histogram. The number of systems in the subpopulation was defined as those having a spin-orbit ratio range of $0.82<P_{\rm orb}/P_{\rm rot}<0.92$. This percentage is significantly less than the one found in L17 of 15\%, but the secondary peak confirms that the subsynchronous EB population observed in L17 is not an artifact specific to the \kepler instrument or survey. We can not conclude from this population percentage alone if the subsynchronous EB population is significant. To assess the significance, we combined the L17 \kepler data with our full \tess confident dataset to create a modeled sample. In this combined sample, $8\%$ of the total population falls within the subsynchronous subpopulation.

\begin{figure}
    \centering
    \includegraphics[scale=0.42]{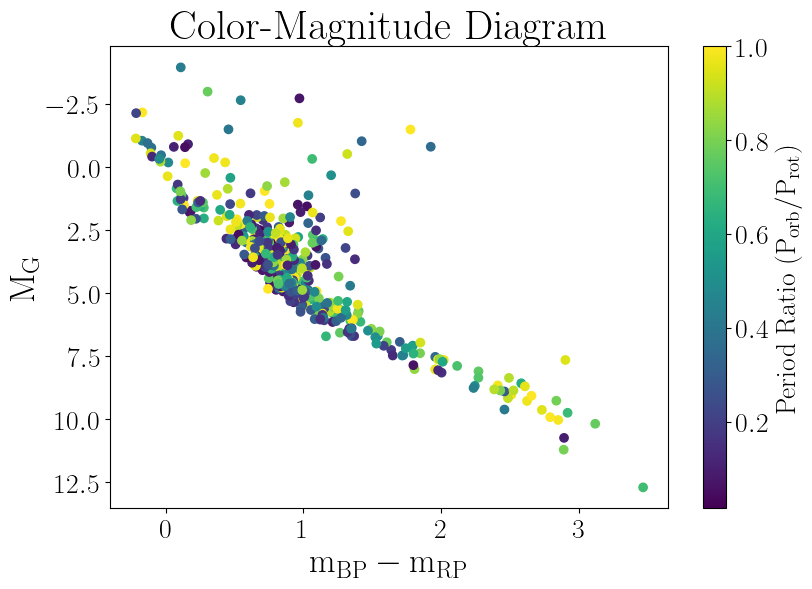}
    \caption{Cross matched EB catalog with \gaia data as a color magnitude diagram, colored by the EB's period ratio.}
    \label{fig:CMD}
\end{figure}
\begin{table}[H]
    \centering
    \begin{tabular}{|c|c|}
    \hline
     Class & $T_{eff}[K]$ \\
     \hline
     O & $\ge$33,000 \\
     B & 10,000–33,000 \\
     A & 7,300–10,000 \\
     F & 6,000–7,300 \\
     G & 5,300–6,000 \\
     K & 3,900–5,300 \\
     M &  2,300–3,900\\
    \hline
    \end{tabular}
    \caption{Effective temperature to stellar classification.}
    \label{tab:star_class}
\end{table}
\subsection{Trends Across Stellar Type and Galactic Location}

To analyze the types of stars present in the subsample, we cross-referenced with TOPCAT \citep{topcat} the RA and DEC of the TICs with the Gaia catalog DR3 \citep{gaiadr3} within 5". Figure \ref{fig:CMD} is a color magnitude diagram of the cross-matched subsample, with a color bar of the spin-orbit ratio. This figure shows how our subsample contains primarily main-sequence stars, with a portion in the subgiant phase. There are some giant and supergiant systems present in the subsample, but there is no clear trend of synchronization with placement on the diagram.

\begin{figure}
    \centering
    \includegraphics[scale=0.45]{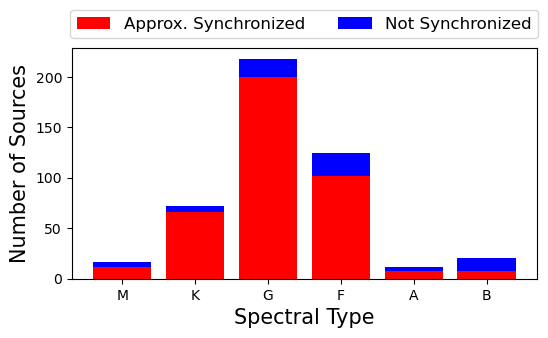}
    \caption{Frequency of spectral types in the \gaia cross-matched eclipsing binary catalog, shown as a stacked bar chart.}
    \label{fig:Star_Classes}
\end{figure}

Confident that we are primarily working with main sequence EBs, we can now map any dominant trends of synchronization across stellar class, using effective temperature and Table \ref{tab:star_class}. Figure \ref{fig:Star_Classes} reveals that a significant portion of our subsample are G and F stars, which is consistent with the distribution of \tess targets \citep{TESS_targets}. There are 116 EBs not included in this figure, due to a missing effective temperature in the cross-matched ID in the \gaia catalog. The systems that are approximately synchronized have $0.8<\frac{P_{orb}}{P_{rot}}<1.2$, and all other ratios are categorized as not synchronized. We conclude that all star types can be synchronized, and there is no dominate synchronized stellar class.

\begin{figure}
    \centering
    \includegraphics[scale=0.45]{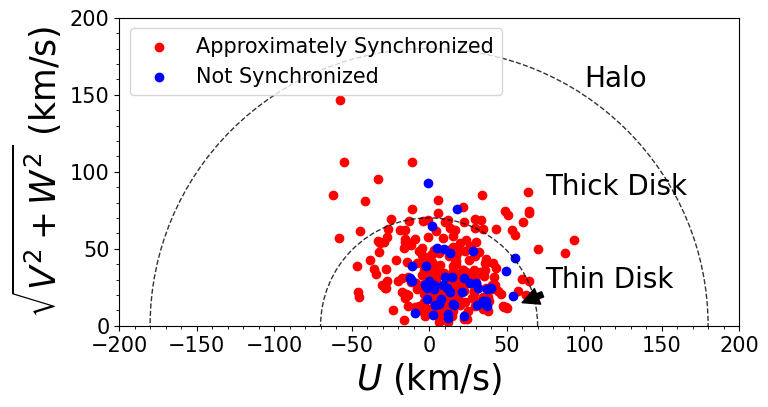}
    \caption{A visualization of our catalog's galactic coordinate positions based on \gaia data. where $V$ represents the velocity (km/s) in the direction of Galactic rotation, $W$ represents the velocity (km/s) toward the North Galactic Pole, and $U$ represents the velocity (km/s) toward the Galactic center. The majority of our catalog is contained in the Milky Way's thin disk, which also houses almost all of the non-synchronous systems.}
    \label{fig:Galac_Loc}
\end{figure}

To cross-check that the EBs are young enough to be main-sequence stars, we check their galactic position. Figure \ref{fig:Galac_Loc} maps the EBs by their galactic kinematics. It is expected that young stars will be in the thin disk and will be older as they migrate into the thick disk, and very old if they are in the halo. Nearly all of the unsynchronized systems in the figure are within the thin disk and thus are young EBs. This result is consistent with previous results that show that older stars are more likely to be synchronized. With an understanding of the types of stars we are examining, we can compare how various rotational period methods influence the spin-orbit ratio. \\
\begin{figure*}
    \subfigure[Single Gaussian posterior corner plot.]{
        \includegraphics[scale=0.7]{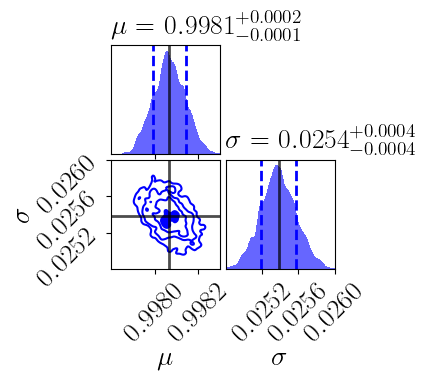}}
    \subfigure[Double Gaussian posterior corner plot.]{
        \includegraphics[scale=0.7]{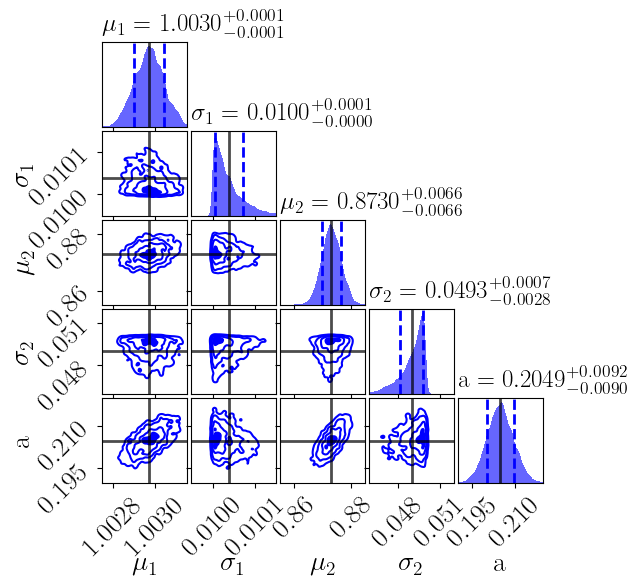}}
    \subfigure{
        \includegraphics[scale=0.7]{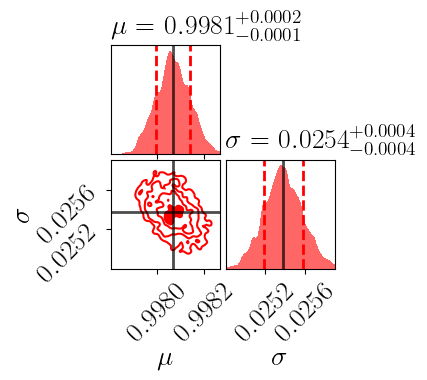}}
    \subfigure{
        \includegraphics[scale=0.7]{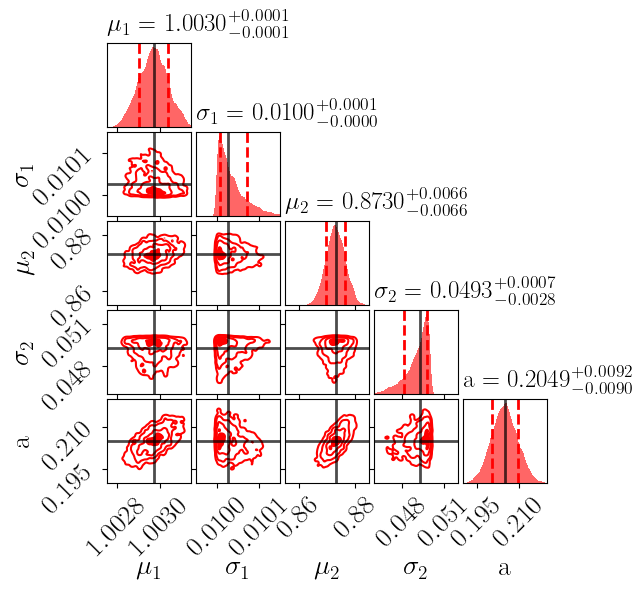}}
    \caption{Corner plots for the weighted mean (blue) and median (red) Gaussian posteriors. The parameter values are shown with black lines. The parameters are $\mu_1$ (primary mean), $\sigma_1$ (primary standard deviation), $\mu_2$ (secondary mean), $\sigma_2$ (secondary standard deviation), a (secondary amplitude). The weighted mean consistently has values within the 5-sigma contour, while the median often does not have values within the 5-sigma bounds.}
    \label{fig:gauss_corners}
\end{figure*}
\begin{figure*}
    \subfigure[Single Lorentz posterior corner plot.]{
        \includegraphics[scale=0.7]{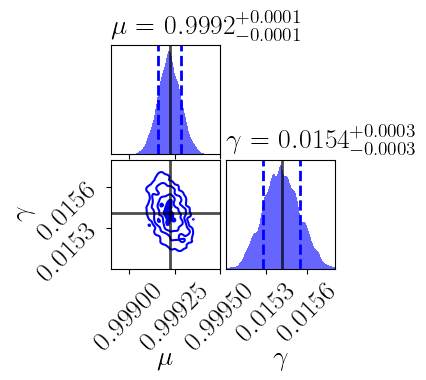}}
    \subfigure[Double Lorentz posterior corner plot.]{
        \includegraphics[scale=0.7]{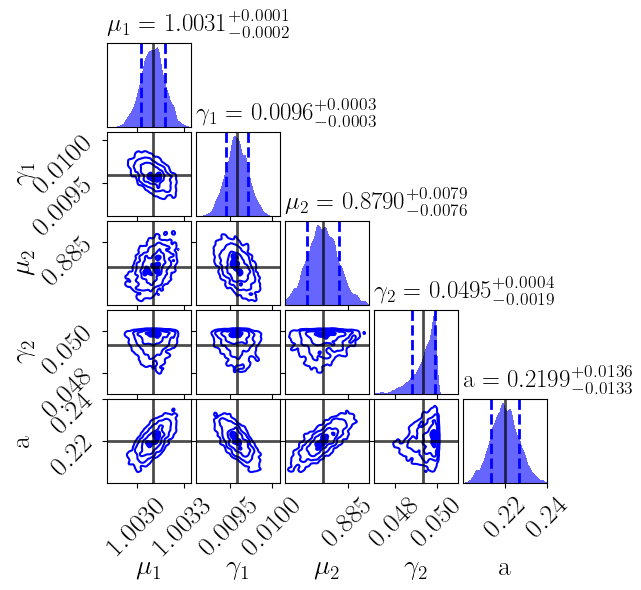}}
    \subfigure{
        \includegraphics[scale=0.7]{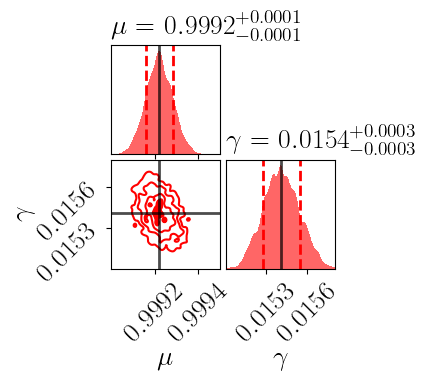}}
    \subfigure{
        \includegraphics[scale=0.7]{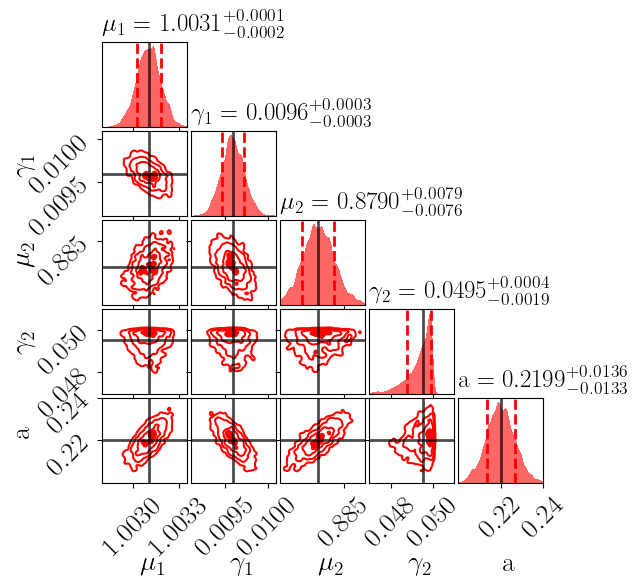}}
    \caption{Corner plots for the weighted mean (blue) and median (red) Lorentz posteriors. The parameter values are shown with black lines. The parameters are $\mu_1$ (primary mean), $\gamma_1$ (primary half width half maximum), $\mu_2$ (secondary mean), $\gamma_2$ (secondary half width half maximum), a (secondary amplitude). The weighted mean consistently has values within the 5-sigma contour, while the median often does not have values within the 5-sigma bounds.}
    \label{fig:cauchy_corners}
\end{figure*}
\subsection{Quantifying the significance of the synchronized population} \label{sec:bayes_factor_results}
\begin{figure*}[htb]
\centering
\begin{turn}{-90}
\begin{minipage}{\textheight}
    \includegraphics[scale=.67]{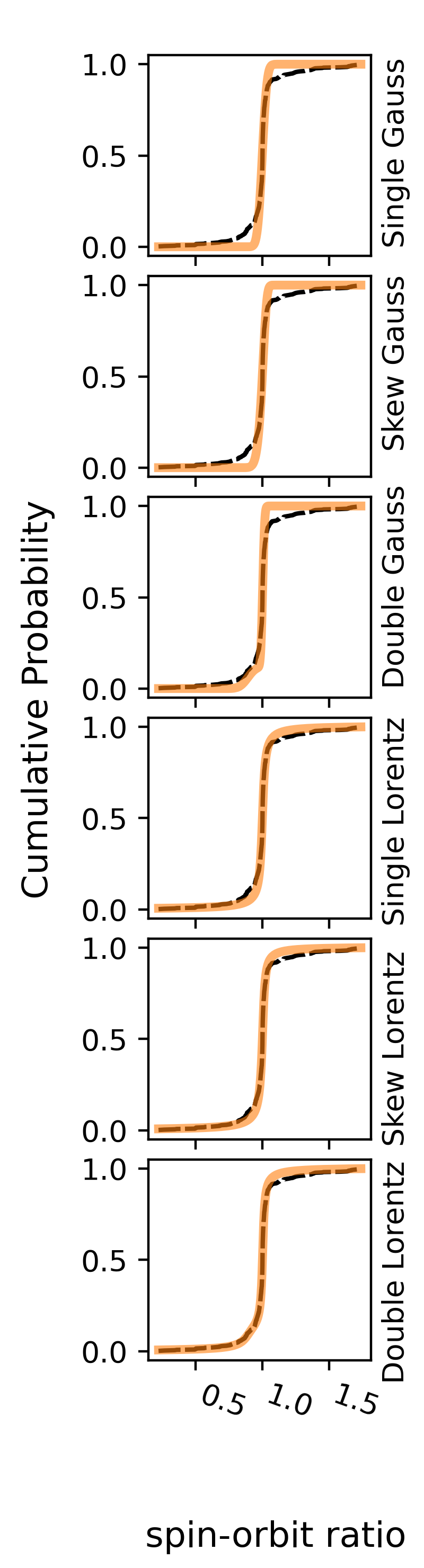}
    \includegraphics[scale=.67]{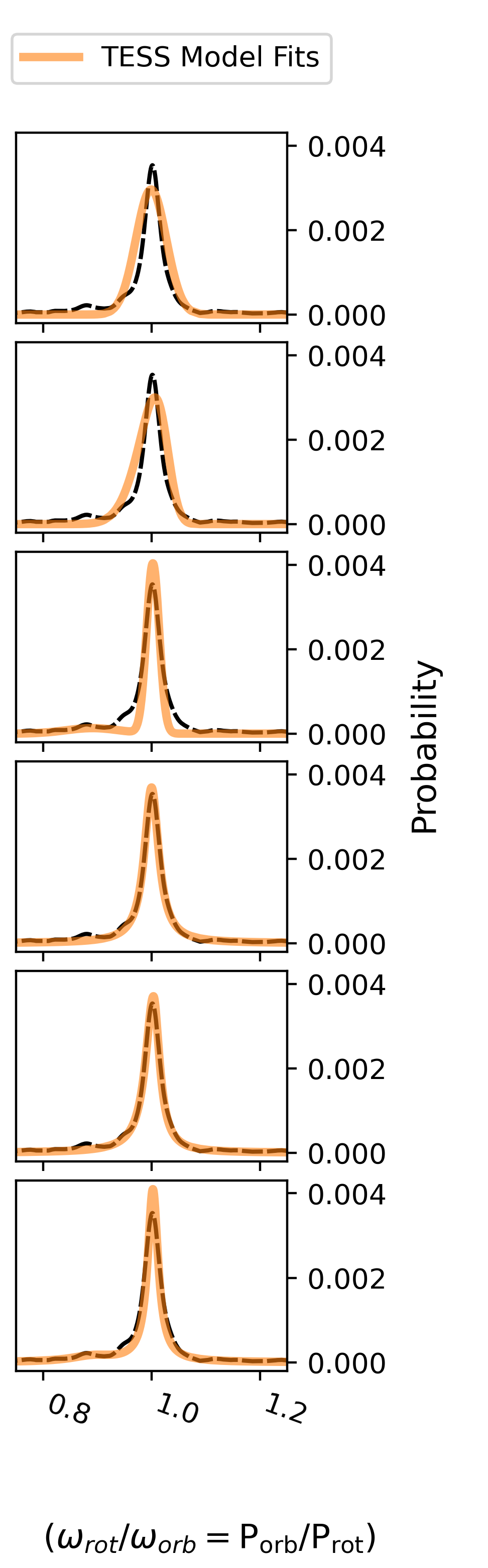}
    \includegraphics[scale=.67]{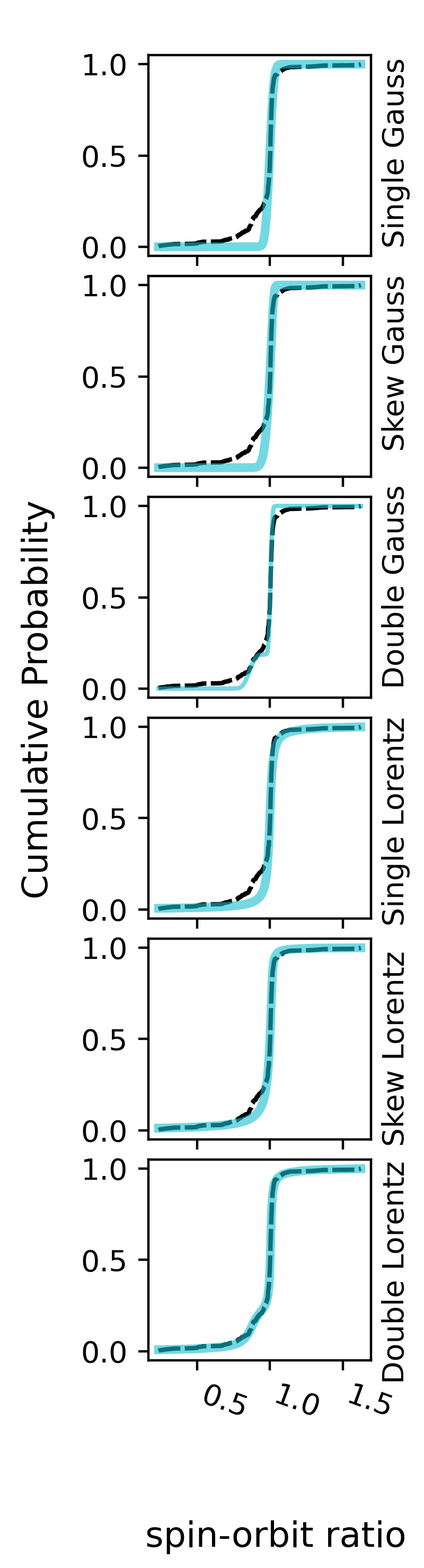}
    \includegraphics[scale=.67]{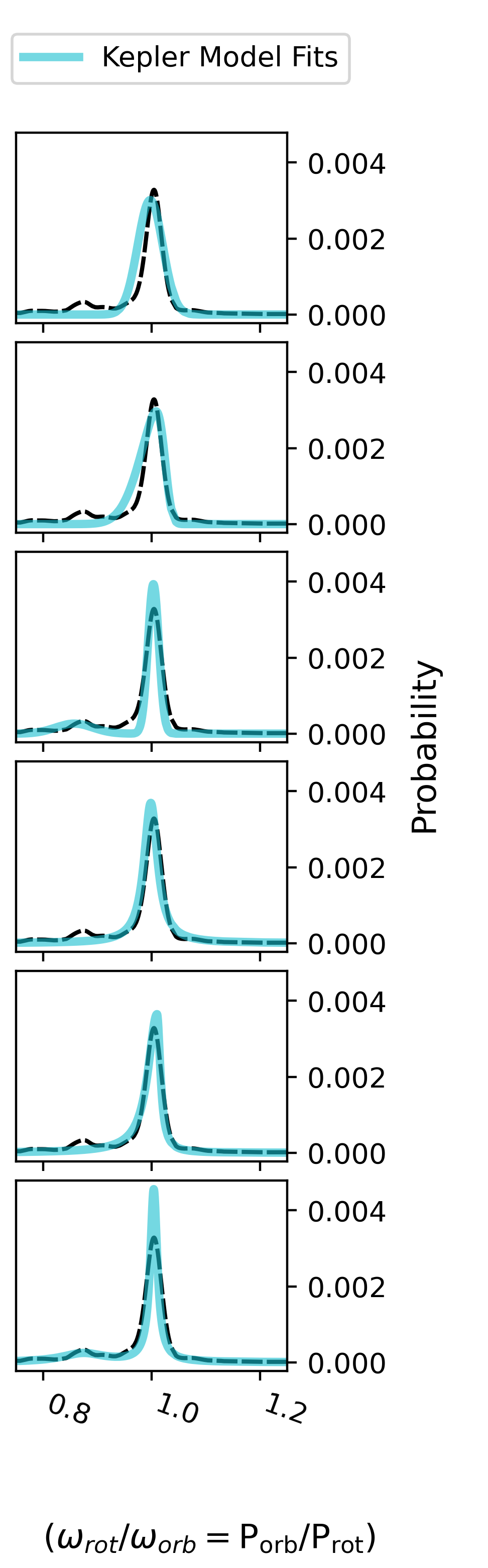}
    \includegraphics[scale=.67]{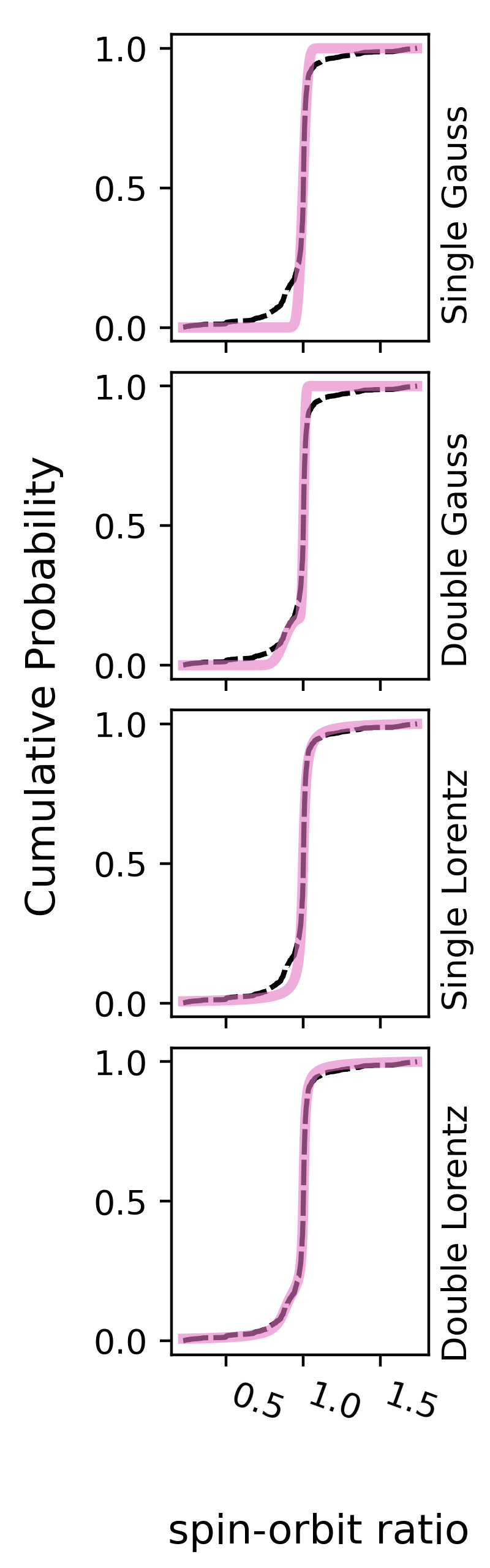}
    \includegraphics[scale=.67]{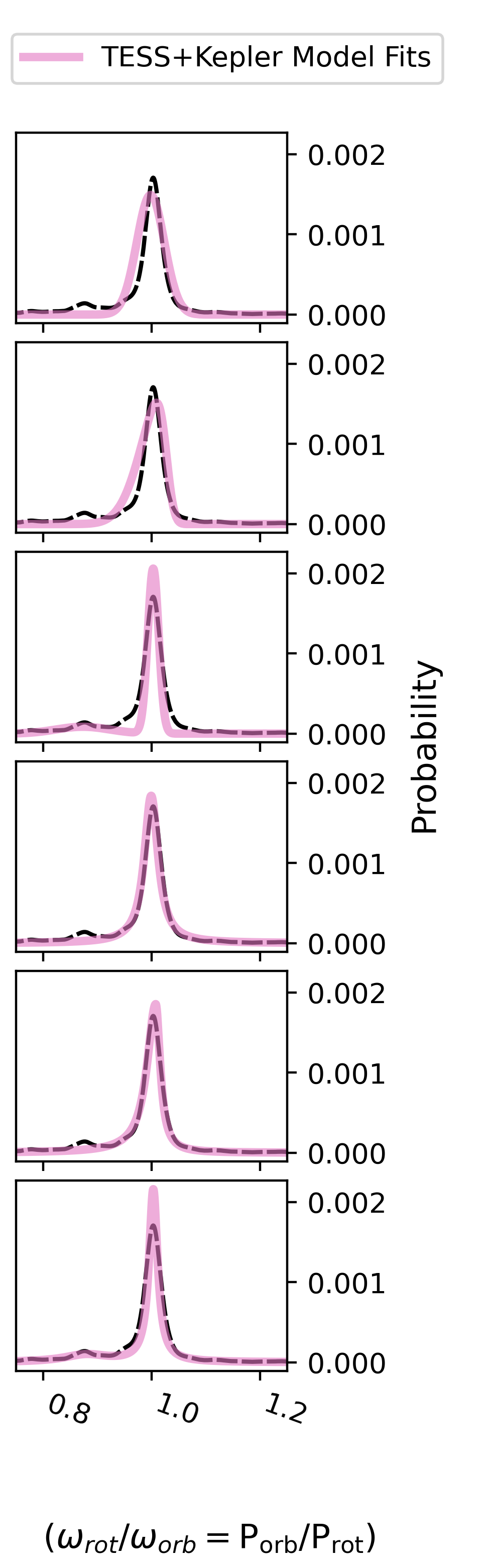} \\
    \caption{Left: Estimation for the cumulative probability density function (CDF) for the combined L17 \kepler and our \tess data sample. The best fit models for the CDF are overlaid for visual comparison. Right: Normalized kernel density estimate of the probability density function (PDF) for the combined L17 \kepler and \tess data sample. The parameters from the best fit models were used to generate PDF functions that are overlaid for visual comparison.}
    \label{fig:CDF/PDF}
\end{minipage}
\end{turn}
\end{figure*}

\begin{figure*}[ht]
    \centering
    \includegraphics[scale=.9]{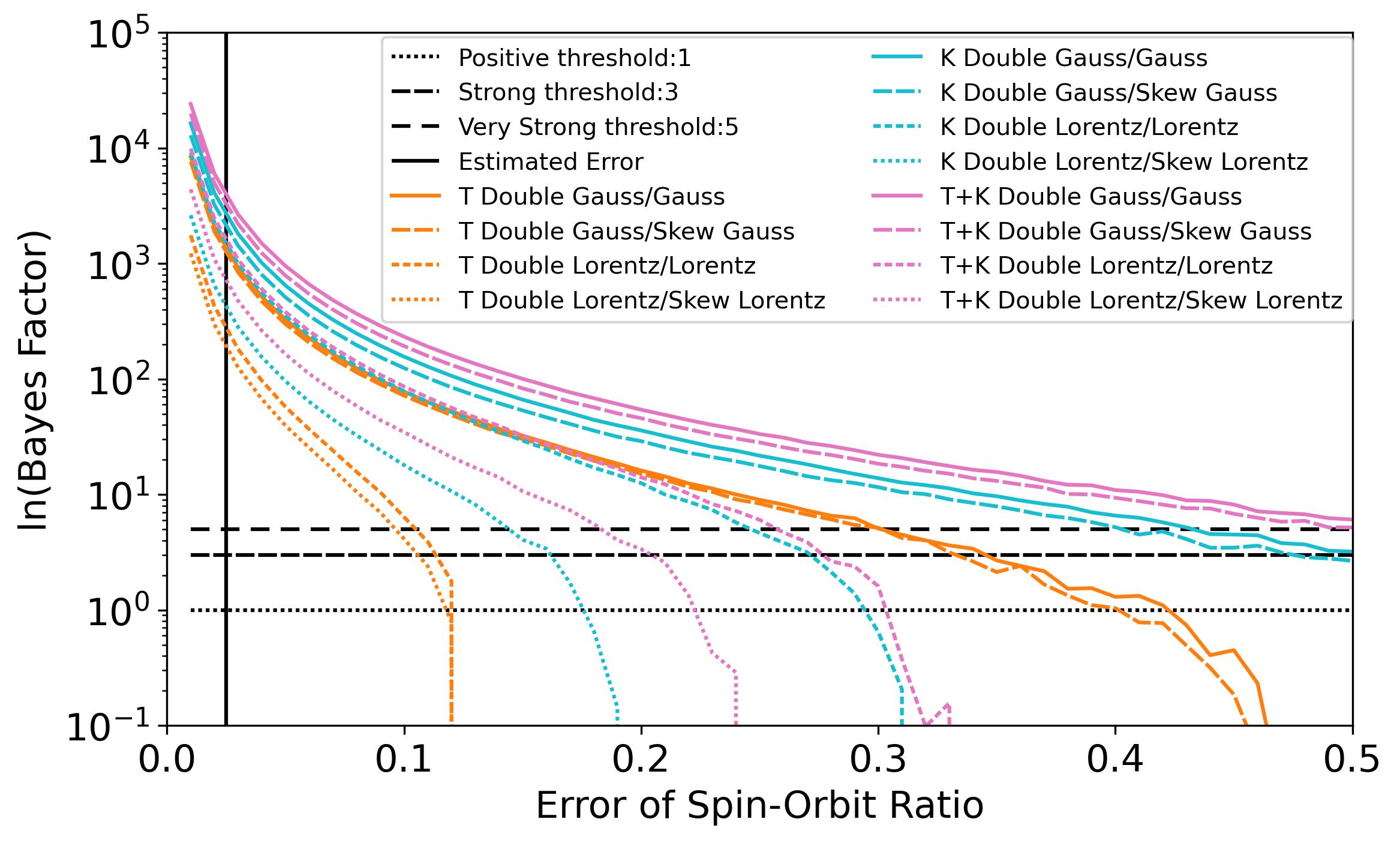}
    \caption{Natural log Bayes factors (BF), representing the natural log of the marginal likelihood ratio of two different models, calculated using a variable uniform error value. The larger the value of the ln BF, the more the first model is favored over the second. Where T is the \tess data, K is the \kepler data, and T+K is the combined data set.}
    \label{fig:BF-error}
\end{figure*}
    
\begin{figure*}[ht]
    \centering
    \includegraphics[scale=0.8]{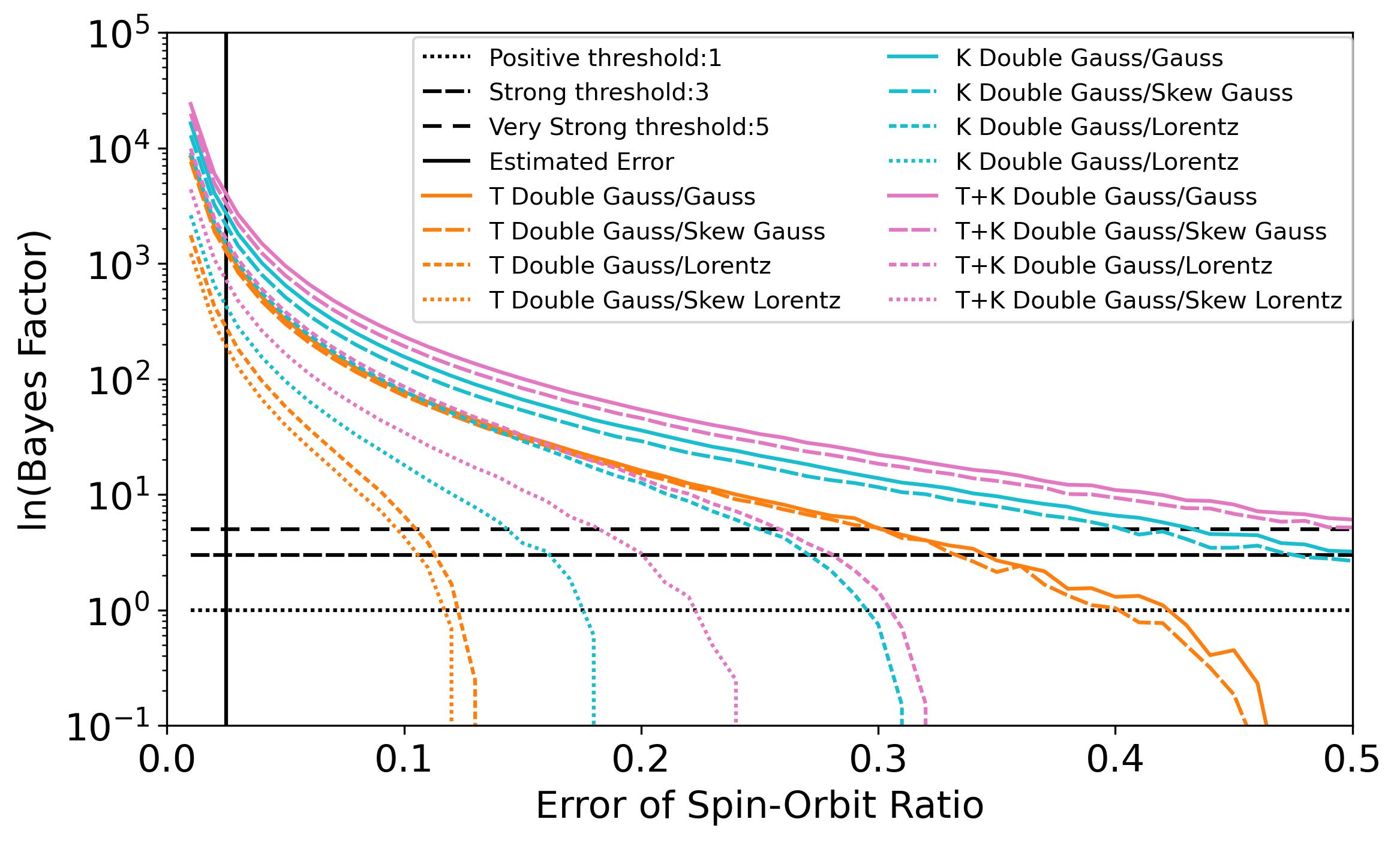}
    \caption{Same dataset as Figure \ref{fig:BF-error}. This plot exclusively compares the double gaussian to the other models, demonstrating that it provides the best fit across all datasets, even at large errors.}
    \label{fig:BF-error-extra-dg}
\end{figure*}

To find the mean and median posterior parameters, we weighted each model's \code{NestedSampler} result. Table \ref{tab:modelparamters} presents the weighted mean and median parameters for all models to the respective data sets. Figures \ref{fig:gauss_corners} and \ref{fig:cauchy_corners} show the corner plots for the combined data set. The corner plots display asymmetry, where the peaks are similar, but the widths of the distributions are slightly skewed. Due to this asymmetry, we employed the median for the remainder of our figures, even though the values of the median and mean values are extremely close. Figure \ref{fig:CDF/PDF} depicts the median posteriors overlaid on both the CDF and normalized PDF across datasets.
\begin{table*}
\begin{center}
\begin{tabular}{ |c|c|c|c|c|c|c|c|c| }
 \hline
 & Model & $\mu_1$ & $\sigma_1|\gamma_1$ & $s_g|s_l$ & $\mu_2$ & $\sigma_2|\gamma_2$ & a \\
 \hline
 \parbox[t]{2mm}{\multirow{4}{*}{\rotatebox[origin=r]{90}{TESS Mean}}} & Single Gauss & 0.9990 & 0.0279 & - & - & -& -\\
 &Skewed Gauss & 1.0277 & 0.0426 & -2.3440 & - & - & -\\
 &Double Gauss &1.0023 & 0.0117 & - &0.8894 & 0.0489 & 0.1368\\ 
 &Single Lorentz & 0.9999 & 0.0159 & - & - & -& -\\ 
 &Skewed Lorentz & 1.0034 & 0.0157 & -0.1309 & - & - & -\\
 &Double Lorentz & 1.0023 & 0.0117 & - & 0.8894 & 0.0489 & 0.1367\\ 
 \hline
 \parbox[t]{2mm}{\multirow{4}{*}{\rotatebox[origin=r]{90}{TESS Median}}} & Single Gauss & 0.9990 & 0.0279 & - & - & -& -\\
 &Skewed Gauss & 1.0278 & 0.0426 & -2.3436 & - & - & -\\
 &Double Gauss &1.0023 & 0.0117 & - & 0.8909 & 0.0493 & 0.1372\\ 
 &Single Lorentz & 0.9999 & 0.0159 & - & - & -& -\\ 
 &Skewed Lorentz & 1.0034 & 0.0157 & -0.1307 & - & - & -\\
 &Double Lorentz & 1.0023 & 0.0117 & - & 0.8909 & 0.0492 & 0.1369\\ 
 \hline
 \parbox[t]{2mm}{\multirow{4}{*}{\rotatebox[origin=r]{90}{Kepler Mean}}} & Single Gauss & 0.9974 & 0.0231 & - & -& -& -\\
 &Skewed Gauss & 1.0246 & 0.0399 & -3.9751 & -& -& -\\
 &Double Gauss &1.0033 & 0.0100 & - & 0.8572 & 0.0346 & 0.2302\\ 
 &Single Lorentz & 0.9985 & 0.0151 & - & - & -& -\\  
 &Skewed Lorentz & 1.0099 & 0.0138 & -0.4583 & - & - & -\\
 &Double Lorentz & 1.0039 & 0.0079 & - & 0.8739 & 0.0477 & 0.3096\\ 
 \hline
 \parbox[t]{2mm}{\multirow{4}{*}{\rotatebox[origin=r]{90}{Kepler Median}}} & Single Gauss & 0.9974 & 0.0231 & - & -& -& -\\
 &Skewed Gauss & 1.0246 & 0.0426 & -2.3436 & - & - & -\\
 &Double Gauss & 1.0033 & 0.0100 & - & 0.8574 & 0.0344 & 0.2301\\ 
 &Single Lorentz & 0.9985 & 0.0151 & - & -& -& -\\ 
 &Skewed Lorentz & 1.0099 & 0.0138 & -0.4583 & - & - & -\\
 &Double Lorentz & 1.0039 & 0.0079 & - & 0.8741 & 0.0483 & 0.3092\\ 
 \hline
 \parbox[t]{2mm}{\multirow{4}{*}{\rotatebox[origin=r]{90}{Combined Mean}}} & Single Gauss & 0.9981& 0.0254 & - & - & -& -\\
 &Skewed Gauss & 1.0272 & 0.0425 & -3.7424 & - & - & -\\
 &Double Gauss &1.0029 & 0.0100 & - & 0.8733&0.0489& 0.2052\\ 
 &Single Lorentz & 0.9992& 0.0154& - & -& -& -\\  
 &Skewed Lorentz & 1.0076 & 0.0147 & -0.3208 & - & - & -\\
 &Double Lorentz & 1.0031& 0.0096& - & 0.8791& 0.0493& 0.2202\\ 
 \hline
 \parbox[t]{2mm}{\multirow{4}{*}{\rotatebox[origin=c]{90}{Combined Median}}} & Single Gauss & 0.9981& 0.0254& - & -& -& -\\
 &Skewed Gauss & 1.0271 & 0.0425 & -3.7428 & - & - & -\\
 &Double Gauss &1.0029&0.0100& - & 0.8732& 0.0492&0.2052\\ 
 &Single Lorentz &0.9992&0.0154& - & -& -& -\\   
 &Skewed Lorentz & 1.0076 & 0.0147 & -0.3207 & - & - & -\\
 &Double Lorentz &1.0031&0.0096& - & 0.8791&0.0493&0.2201\\ 
 \hline
\end{tabular}
\caption{Weighted mean and weighted median parameters of each posterior. The parameters are $\mu_1$ (primary mean), $\sigma_1|\gamma_1$ (primary standard deviation$|$half width half maximum), $\mu_2$ (secondary mean), $\sigma_2|\gamma_2$ (secondary standard deviation$|$half width half maximum), $s_g|s_l$ (Gaussian skew$|$Lorentzian skew), a (secondary amplitude). The primary amplitude is always 1, so it is not a parameter.}
\label{tab:modelparamters}
\end{center}
\end{table*}

To compare these best-fit models, we calculated the natural log of the Bayes factors (BF) for all models. Specifically, we examined how the Bayes factors depend on the spin-orbit ratio error for two different model comparisons: a double Gaussian versus a single Gaussian, and a double Lorentzian versus a single Lorentzian. These comparisons were selected because our primary interest lies in assessing the significance of the secondary population, rather than identifying the optimal distribution type.

The significance of the natural log Bayes factor ($\ln\mathcal{B}$) falls into the following ranges \citep{doi:10.1080/01621459.1995.10476572}: $\ln\mathcal{B} < 0$ supports $Z_2$, $0 < \ln\mathcal{B} < 1$ barely supports $Z_1$, $1 < \ln\mathcal{B} < 3$ positively supports $Z_1$, $3 < \ln\mathcal{B} < 5$ strongly supports $Z_1$, and $\ln\mathcal{B} > 5$ very strongly supports $Z_1$. Table \ref{tab:Ln Bayes factor} presents the calculated natural log Bayes factors across posteriors, demonstrating that a double-peaked distribution, is decisively the best fit across data sets. For the \tess data, a double distribution is the best fit, with a small favorability towards the double Lorentz. However, a double Lorentzian decisively provides the best fit for both the \kepler and combined data sets.
\begin{table*}
\begin{center}
\begin{tabular}{ |c|c||c|c|c|c|c|c| }
 \hline
 & Model & Single Gauss & Skewed Gauss & Double Gauss & Single Lorentz & Skewed Lorentz & Double Lorentz \\
 \hline
 \hline
 \parbox[t]{2mm}{\multirow{4}{*}{\rotatebox[origin=r]{90}{\tess}}}&Single Gauss & - & 118 & 1330 & 1062 & 1145 &1330 \\
 &Skewed Gauss & -118 & - & 1212 & 944 & 1026 & 1212 \\
 &Double Gauss & -1330 & -1212 & - & -267 & -185 & 0.24 \\ 
 &Single Lorentz & -1062 & -944 & 267 & - & 82 & 267 \\ 
 &Skewed Lorentz & -1145 & -1026 & 185 & -82 & - & 186 \\
 &Double Lorentz & -1330 & -1212 & -0.24 & -267 & -186 & - \\ 
 \hline
\parbox[t]{2mm}{\multirow{4}{*}{\rotatebox[origin=r]{90}{\kepler}}}&Single Gauss & - & 550 & 2607 & 1162 & 2198 & 2680 \\
 &Skewed Gauss & -550 & - & 2057 & 612 & 1647 & 2130 \\
 &Double Gauss & -2607 & -2057 & - & -1444 & -410 & 73 \\ 
 &Single Lorentz & -1162 & -612 & 1444 & - & 1035 & 1572 \\ 
 &Skewed Lorentz & -2198 & -1647 & 410 & -1035 & - & 483 \\ 
 &Double Lorentz & -2680 & -2130 & -73 & -1572 & -483 & - \\ 
 \hline
  \parbox[t]{2mm}{\multirow{4}{*}{\rotatebox[origin=r]{90}{Combined}}}&Single Gauss & - & 671 & 3838 & 2272 & 3148 & 3844\\
 &Skewed Gauss & -671 & - & 3167 & 1601 & 2476 & 3173\\
 &Double Gauss & -3838 & -3167 & - & -1566 & -690 & 6.2\\ 
 &Single Lorentz & -2272 & -1601 & 1566 & - & 875 & 1572\\ 
 &Skewed Lorentz & -3148 & -2476 & 690 & -875 & - & 697\\
 &Double Lorentz & -3844 & -3173 & -6.2 & -1572 & -697 & -\\ 
 \hline
\end{tabular}
\caption{Natural log Bayes factors comparing the evidence between each pair of model, from a set of four different candidate models: single Gaussian, double Gaussian, single Lorentzian, and double Lorentzian. As given in Equation \ref{eq:BF}, the natural log Bayes factor is $\ln\mathcal{B} = \ln(\frac{Z_1}{Z_2}) = \ln{Z_1}-\ln{Z_2}$, where $Z_1$ is the model listed in the column, and $Z_2$ is the model listed in the corresponding row. }
\label{tab:Ln Bayes factor}
\end{center}
\end{table*}

Given that our results are contingent on a rough estimate of the uncertainty of the spin-orbit ratio, we computed various natural log Bayes factors across a range of errors to assess if it would alter our conclusions. Using the same process as described in Section \ref{bayesian}, we tested varying the spin-orbit ratio uncertainty between 0.01 and 0.5, in the likelihood calculations. Figure \ref{fig:BF-error}  shows how our confidence in model selection depends on the typical error value chosen across data points. The dashed horizontal black lines indicate the significance levels of the Bayes factor and the solid black vertical line shows our fiducial error value of $P_{orb}/P_{rot}$ = 0.025. This figure illustrates that, even with large error values, a double-peaked distribution remains decisively superior to a single-peaked distribution, affirming the reality of the subpopulation regardless of the true error values. See Figure \ref{fig:BF-error-extra-dg} for visualization that a double Gaussian is decisively the best fit. \\


\clearpage
\section{Conclusion} \label{sec:conclusion}

We systematically examined EB light curves from TESS via Lomb-Scargle periodograms, the autocorrelation function, phase dispersion minimization and by-eye methods ($\S$ 4.3) to constrain rotational periods as a function of mass, orbital period, age, and galactic location. We provide our data set in the supplementary material. We recovered previous trends that are consistent with tidal theory, such as a preponderance of synchronized binaries with orbital periods less than 10 days, a higher probability of synchronization for older binaries, and tentative detections of rotational periods in higher order spin-orbit resonances for larger orbital periods. Crucially, we also found the population of binaries with spin-orbit ratios of 7:8 that L17 found in Kepler data. We performed Bayes factor tests on the spin-orbit ratio distribution and found that these 7:8 rotators cannot be attributed to noise and are distinct from the synchronous rotators ($\S$ 4.4). We therefore conclude that $\sim$10\% of short-period EBs rotate very closely to 7/8ths of their orbital period instead of synchronously.

The origins of these 7:8 rotators, however, remains a mystery. As we generally recover the spin-orbit ratio from L17, the F19 results are applicable to our study as well. That experiment coupled tidal + stellar evolution to show that EBs with orbital periods between 3 and 10 days can spend significant time near the 7:8 ratio, but with significantly more spread than in the observational data. L17 also suggested differential rotation could generate the observations but that hypothesis remains untested. Other possibilities exist, such as a spin-orbit resonance resulting from a triaxial torque or dynamical tides, but they also remain untested. Our results here motivate more theoretical work on the subject since we have recovered this overdensity with a different instrument and mathematical period-finding algorithms, as well as different people vetting the data by eye.

Nonetheless, additional work on measuring rotation periods is still warranted. Our methods do not generate robust measurement uncertainties on the individual rotational periods, which in turn limits statistical methods of confirming the presence of the 7:8 rotators.  A more accurate method would be to model the stellar variability using a Gaussian process \cite[\textit{e.g.}][]{angus_inferring_2018,angus_exploring_2020,gordon_stellar_2021}, which would determine posteriors of the rotation period for each individual source. However, this analysis would require implementing a whole different, more computationally expensive model and is beyond the scope of this paper.

In summary, we have systematically mined \tess data to identify and characterize EBs. While our focus was on confirming the 7:8 spin-orbit resonance population, this catalog also enables future studies of circumbinary exoplanets, differential rotation, eccentricity, magnetic braking, pulsators, tidal synchronization, and tidal torque. The methodology presented here can be extended to future analysis of EB light curves, advancing our understanding of stellar properties and evolution.\\


This paper includes data collected with the TESS mission, obtained from the MAST data archive at the Space Telescope Science Institute (STScI). Funding for the TESS mission is provided by the NASA Explorer Program. STScI is operated by the Association of Universities for Research in Astronomy, Inc., under NASA contract NAS 5–26555.

This paper includes data collected by the Kepler mission and obtained from the MAST data archive at the Space Telescope Science Institute (STScI). Funding for the Kepler mission is provided by the NASA Science Mission Directorate. STScI is operated by the Association of Universities for Research in Astronomy, Inc., under NASA contract NAS 5–26555.

This work has made use of data from the European Space Agency (ESA) mission
{\it Gaia} (\url{https://www.cosmos.esa.int/gaia}), processed by the {\it Gaia}
Data Processing and Analysis Consortium (DPAC,
\url{https://www.cosmos.esa.int/web/gaia/dpac/consortium}). Funding for the DPAC
has been provided by national institutions, in particular the institutions
participating in the {\it Gaia} Multilateral Agreement.

This work made use of ChatGPT, a large language model developed by OpenAI, to assist with rewording sentences and improving the flow of the text. ChatGPT was accessed via the ChatGPT web interface (https://chat.openai.com).

\software{\code{astropy} \citep{astropy_collaboration_astropy_2013,astropy_collaboration_astropy_2018,astropy_2022ApJ},  
          \code{cesium} \citep{cesium}, \code{dynesty} \citep{dynesty_2020},
          \code{exoplanet} \citep{foreman-mackey_exoplanet_2021},
          \code{lightkurve} \citep{lightkurve}
          }

\clearpage
\bibliography{ref.bib}
\bibliographystyle{aasjournal}

\end{document}